\documentclass[10pt]{iopart}
\usepackage{iopams}  
\usepackage[T1]{fontenc}
\usepackage[utf8]{inputenc}
\expandafter\let\csname equation*\endcsname\relax
\expandafter\let\csname endequation*\endcsname\relax
\usepackage{amsmath}
\usepackage{graphicx}
\usepackage{amsxtra}
\usepackage{empheq}
\usepackage{amsfonts}
\usepackage{amssymb}
\usepackage{amsthm}
\usepackage{enumitem}
\usepackage{tcolorbox}
\usepackage{url}
\usepackage{color,hyperref}
\usepackage{tikz-cd}
\usepackage{epstopdf}
\usepackage{titlesec}
\usepackage{color}
\usetikzlibrary{calc,arrows,decorations.pathmorphing}
\hypersetup{colorlinks,breaklinks,
             linkcolor=blue,urlcolor=blue,
           anchorcolor=blue,citecolor=blue}
\usepackage{cite}
\usepackage[margin=3cm]{geometry}

\newcommand{\R}{\mathbb{R}}

\newcommand{\E}{\mathbb{E}}

\newcommand{\vb}[1]{\mathbf{#1}}

\newcommand{\lp}{\left(}
\newcommand{\rp}{\right)}
\newcommand{\lb}{\left[}
\newcommand{\rb}{\right]}

\usepackage{titlesec}

\makeatletter
\newcommand{\mainmatter}{%
  \setcounter{footnote}{1}%
  \patchcmd{\@makefntext}{\fnsymbol}{\arabic}{}{}%
  \patchcmd{\@thefnmark}{\fnsymbol}{\arabic}{}{}%
  \def\@makefnmark{\textsuperscript{\arabic{footnote}}}%
}
\makeatother

\newtcolorbox[auto counter,number within=section]{pabox}[2][]{%
colback=black!5!white,colframe=black!60!white,fonttitle=\bfseries,
title=Method~\thetcbcounter: #2,#1}

\newcommand{\Var}{\operatorname{Var}}

\newcommand{\wt}{\widetilde}
\newcommand{\wh}{\widehat}

\usepackage[labelfont=bf,font=small,skip=3pt]{caption}
\titleformat*{\subsection}{\normalfont}

\begin{document}

\title[]{Accelerating the estimation of energetic particle confinement statistics in stellarators using multifidelity Monte Carlo}

\author{Frederick Law\footnote[1]{To
whom correspondence should be addressed (law@cims.nyu.edu)}, Antoine Cerfon, Benjamin Peherstorfer}

\address{Courant Institute of Mathematics, New York University}
\vspace{10pt}

\begin{abstract}
In the design of stellarators, energetic particle confinement is a critical point of concern which remains challenging to study from a numerical point of view. Standard Monte Carlo analyses are highly expensive because a large number of particle trajectories need to be integrated over long time scales, and small time steps must be taken to accurately capture the features of the wide variety of trajectories. Even when they are based on guiding center trajectories, as opposed to full-orbit trajectories, these standard Monte Carlo studies are too expensive to be included in most stellarator optimization codes. We present the first multifidelity Monte Carlo scheme for accelerating the estimation of energetic particle confinement in stellarators. Our approach relies on a two-level hierarchy, in which a guiding center model serves as the high-fidelity model, and a data-driven linear interpolant is leveraged as the low-fidelity surrogate model. We apply multifidelity Monte Carlo to the study of energetic particle confinement in a 4-period quasi-helically symmetric stellarator, assessing various metrics of confinement. Stemming from the very high computational efficiency of our surrogate model as well as its sufficient correlation to the high-fidelity model, we obtain speedups of up to 10 with multifidelity Monte Carlo compared to standard Monte Carlo.
\end{abstract}

\mainmatter

\submitto{\NF}


\section{Introduction} \label{section:intro}
In generic non-axisymmetric magnetic configurations, collisionless particle orbits are not guaranteed to be confined \cite{Helander_2014}. Energetic particle confinement therefore is a key point of concern for the design of tokamaks which have significant toroidal ripple and of stellarators \cite{Lotz_1992,Grieger:1992,Beidler2001,Najmabadi_2008,Bunno:2013,Drevlak_2014,Bader:2019,Henneberg_2019}. In order to design configurations which maximize energetic particle confinement, one must identify metrics which accurately capture the level and quality of confinement, and are at the same time as easy and as inexpensive to include in design optimization codes as possible. 

One can distinguish two distinct approaches to estimating energetic particle confinement. The first approach may be viewed as a direct approach, or Monte Carlo approach. One launches an ensemble of energetic particles, follows their individual orbits---usually their drift orbits---and then directly calculates the fraction of particles which have been lost and the time of flight of particles before being lost \cite{Grieger:1992,Reiman_1999,Ku_2006,Spong_2011,Nemov_2014,Pfefferle_2016,Henneberg_2019,Albert:2020,Bader2020new,bader2021energetic}. On the one hand, this method can be considered as physically accurate, since it directly captures the effects on confinement of the wide variety of orbits energetic particles follow in nonaxisymmetric magnetic configurations. On the other hand, the convergence of the Monte Carlo estimator as a function of the number of particles launched is slow, requiring a large ensemble of particle trajectories for accurate estimations. This means that for guiding center calculations this first method is more time-consuming than most other calculations involved in the design optimization of nonaxisymmetric equilibria, and prohibitively expensive for full-orbit trajectories. Furthermore, this method does not lend itself easily to the computation of derivatives of the metric with respect to magnetic field variations, which implies that it cannot be easily implemented in gradient-based optimization schemes.

The second approach used for estimating the quality of energetic particle confinement is deterministic and relies on proxy functions with closed analytic forms, which can be directly computed from the output of a nonaxisymmetric equilibrium solver, and which have empirically been found to often correlate well with energetic particle confinement. One may for example make the approximation that good energetic particle confinement will be a direct consequence of good low collisionality neoclassical transport properties, and thus characterize the confinement performance through the evaluation of the effective ripple $\epsilon_{\mathrm{eff}}$ first introduced by Nemov \cite{Nemov_1999,Spong_2015}. In order to target energetic particle confinement more directly, one may instead rely on another metric recently proposed by Nemov, called the $\Gamma_{c}$ metric \cite{Nemov_2008,Bader:2019}. $\Gamma_{c}$ is better suited for this purpose because it applies to fully collisionless orbits and unlike $\epsilon_{\mathrm{eff}}$ it is also strongly influenced by particles close to the trapped/passing boundary, which often are the dominant contribution to energetic particle losses \cite{Bader:2019,Bader2020new}. The benefits of these proxy quantities for energetic particle confinement are three folds. First, they are significantly less expensive computationally than Monte Carlo based solely on expensive, high-fidelity simulations. Second, their closed analytic forms, which were derived from physics principles, enable easier interpretation of the numerical results and identification of trends. Third, it may be easier to compute with high accuracy their derivatives with respect to optimization parameters, as was recently demonstrated \cite{Paul_2020}.

Despite the advantages of the deterministic approach, its application may remain limited to preliminary optimization studies, because the currently used deterministic functions have known limitations in their predictive capabilities \cite{Albert:2020}. In the study of a vacuum quasi-helically symmetric configuration, it was for example recently found that energetic particle confinement could be anti-correlated with $\epsilon_{\mathrm{eff}}$ \cite{Bader:2019}: energetic particle confinement was improved while at the same time the $\epsilon_{\mathrm{eff}}$ metric increased. The $\Gamma_{c}$ metric was found to be reliable in that same study \cite{Bader:2019}, and a strong correlation between alpha particle energy loss and $\Gamma_{c}$ in the outer half of the plasma was recently highlighted for a wide variety of stellarator configurations \cite{bader2021energetic}. However, that correlation is not perfect. Stellarators with significantly different $\Gamma_{c}$ values can have comparable levels of energy loss, and stellarators with comparable $\Gamma_{c}$ values can have significantly different levels of energy loss. It therefore appears that for the goal of implementing reliable and efficient energetic particle confinement calculations in multi-objective optimization codes, it is necessary to include high-fidelity simulations with lower costs than a direct Monte Carlo approach. This is precisely the purpose of this article:  we apply the multifidelity Monte Carlo (MFMC) method \cite{Ng:2014,Peherstorfer:2016,Peherstorfer:2018} to reduce the variance of the Monte Carlo estimator for energetic particle confinement, and thus reduce the computational cost of such an estimator.

As we will explain in more detail in the main text, the MFMC method is a strategy for variance reduction of Monte Carlo estimators based on  control variates; see, e.g., \cite{MCMethods,nelson_control_1987}.
In MFMC, the control variates are given by low-fidelity models, i.e. approximations of the high-fidelity physics model one is interested in, which are carefully designed to satisfy two key criteria: they have a high correlation with the high-fidelity model, and they are significantly less expensive to evaluate than the high-fidelity model. These criteria are required to obtain variance reduction as shown in \cite{Ng:2014,Peherstorfer:2016,Peherstorfer:2018}. The MFMC approach is well suited to situations in which scientists rely on the Monte Carlo method to propagate the uncertainty associated with the initial conditions and with parameters defining the model, as is precisely the case for the confinement of energetic particles in nonaxisymmetric magnetic configurations. The method and some of its close variants have recently been applied for the first time to kinetic plasma models \cite{DiMarco2019,DiMarco2020,konrad2021}. In \cite{DiMarco2019,DiMarco2020}, the authors demonstrate that physics based reduced models can be effectively used for variance reduction of the Monte Carlo estimator for the solution of the Boltzmann equation subject to uncertainty, in highly collisional plasmas not subject to a far-field force field. In contrast, in \cite{konrad2021}, the authors rely on data-driven low-fidelity models to reduce the variance of their Monte Carlo estimators for quantifying the uncertainty in kinetic microturbulence. To the best of our knowledge, our article represents the first time the MFMC method is applied to the study of energetic particle confinement in stellarators, and stellarator performance in general. As in \cite{konrad2021}, we rely on a data-driven low-fidelity model for MFMC based variance reduction, since traditional sources of low-fidelity models are inappropriate for our high-fidelity model. For example, our low-fidelity model cannot just utilize simplified physics because our high-fidelity model already is a reduced physics model, i.e. collisionless guiding center trajectories, and we are not aware of simpler physics based models which maintain a reasonably high correlation with the high-fidelity model during the entire energetic particle trajectory. We report that despite the large variety of particle orbits depending on the initial conditions, and despite the chaotic nature of the dynamics, our relatively simple data-driven model---which can be trained with standard tools available in common scientific software packages in Python and Matlab---can lead to significant variance reduction in the MFMC framework, and corresponding speed up.

The structure of the article is as follows. In Section \ref{section:dynamics}, we present our guiding center model for collisionless energetic particle trajectories, and briefly discuss why numerical simulations based on this model are intrinsically expensive. In Section \ref{section:UQ-setup} we give a brief review of the standard Monte Carlo approach for estimating energetic particle confinement, which we contrast with our new MFMC approach in Section \ref{section:MFMC}. We describe the set up for our numerical tests and demonstrate the speed-up obtained with our MFMC method in these tests in Section \ref{section:numerics}. We summarize our work in Section \ref{section:conclusion}, where we also identify some of its limitations, and suggest corresponding directions for future work.

\section{Energetic particle dynamics} \label{section:dynamics}
The main purpose of our article is to demonstrate in a realistic setting the potential of the MFMC method to reduce the variance of the commonly used Monte Carlo estimators for energetic particle confinement, and thus to speed up the evaluation of these estimators for optimization applications. For simplicity, we will consider Monte Carlo estimators based on collisionless orbits, as is still often done in stellarator optimization studies \cite{Drevlak_2018,Bader:2019,ColeCollisionlessXGC,Henneberg_2019,Bader2020new}. Our MFMC approach will be directly applicable to Monte Carlo estimators based on single-particle orbits including collisions as well  \cite{Ku_2006,Pfefferle_2016,Henneberg_2019}. In this section, we begin by detailing the equations of motion we solve to compute the trajectories of energetic particles. We then highlight certain features of the model which underscore the expensive nature behind simulating particle trajectories.

\subsection{Guiding center trajectories}
We consider the collisionless dynamics of 3.5 MeV alpha particles as the byproduct of deuterium-tritium fusion in a stellarator magnetic field. We assume the magnetic configuration has nested flux surfaces which enables us to work in flux coordinates: $(s,\theta, \zeta)$ are the coordinates of our flux coordinate system, where $s \in [0,1]$ is the normalized flux surface label, $\theta$ is a poloidal angle, and $\zeta$ is a toroidal angle. In this coordinate system, we identify $s = 0$ with the magnetic axis and $s = 1$ with the last closed flux surface.

We model the dynamics of energetic particles using the \textit{guiding center equations} \cite{Littlejohn:1983,Landreman:2018,Boozer:1980,Cary:2009} which arise from gyro-averaging the full-orbit equations given by the Lorentz force. Our multi-fidelity Monte Carlo approach is in principle also applicable to Monte Carlo estimators based on full orbits of energetic particles. However, simulating such orbits on the desired long $\alpha$-particle slowing down time \cite{spitzer1956physics,Lotz_1992} remains prohibitively expensive for optimization studies on existing computing architectures. To highlight the immediate relevance of our work for tokamak and stellarator optimization studies, we thus chose to focus on drift orbits. The guiding center equations can be expressed in terms of a 4-dimensional system of coupled ordinary differential equations (ODE) for $(\vb{r},v_{||})$, where $\vb{r}$ is the position of the guiding center and $v_{||}$ is the parallel velocity of the energetic particle. In flux coordinates $(s,\theta,\zeta)$, the guiding center equations take the form
\begin{align}
    \label{guiding center}
    \frac{ds}{dt} =& \frac{1}{\Omega \sqrt{g}} \lb - \mu \frac{B_{\zeta}}{B} \frac{\partial B}{\partial \theta} + \mu \frac{B_{\theta}}{B} \frac{\partial B}{\partial \zeta} + \frac{v_{||}^{2}}{B} \frac{\partial B_{\zeta}}{\partial \theta}- \frac{v_{||}^{2} B_{\zeta}}{B^{2}} \frac{\partial B}{\partial \theta} - \frac{v_{||}^{2}}{B} \frac{\partial B_{\theta}}{\partial \zeta}  + \frac{v_{||}^{2} B_{\theta}}{B^{2}} \frac{\partial B}{\partial \zeta}\rb\\
    \frac{d\theta}{dt} =& \frac{v_{||} B^{\theta}}{B} + \frac{1}{\Omega \sqrt{g}} \lb \mu \frac{B_{\zeta}}{B} \frac{\partial B}{\partial s} - \mu \frac{B_{s}}{B} \frac{\partial B}{\partial \zeta} - \frac{v_{||}^{2}}{B} \frac{\partial B_{\zeta}}{\partial s} + \frac{v_{||}^{2} B_{\zeta}}{B^{2}} \frac{\partial B}{\partial s} + \frac{v_{||}^{2}}{B} \frac{\partial B_{s}}{\partial \zeta} - \frac{v_{||}^{2} B_{s}}{B^{2}} \frac{\partial B}{\partial \zeta}\rb\\
    \frac{d\zeta}{dt} =& \frac{v_{||}^{2} B^{\zeta}}{B} + \frac{1}{\Omega \sqrt{g}} \lb - \mu \frac{B_{\theta}}{B} \frac{\partial B}{\partial s} + \mu \frac{B_{s}}{B} \frac{\partial B}{\partial \theta} + \frac{v_{||}^{2}}{B} \frac{\partial B_{\theta}}{\partial s} - \frac{v_{||}^{2} B_{\theta}}{B^{2}} \frac{\partial B}{\partial s} - \frac{v_{||}^{2}}{B} \frac{\partial B_{s}}{\partial \theta}  + \frac{v_{||}^{2} B_{s}}{B^{2}} \frac{\partial B}{\partial \theta}\rb\\
    \frac{dv_{||}}{dt} =& - \frac{\mu v_{||}}{B \Omega \sqrt{g}} \lb \frac{\partial B}{\partial s} \frac{\partial B_{\zeta}}{\partial \theta} - \frac{\partial B}{\partial s} \frac{\partial B_{\theta}}{\partial \zeta} - \frac{\partial B}{\partial \theta} \frac{\partial B_{\zeta}}{\partial s} + \frac{\partial B}{\partial \theta} \frac{\partial B_{s}}{\partial \zeta} +  \frac{\partial B}{\partial \zeta} \frac{\partial B_{\theta}}{\partial s} - \frac{\partial B}{\partial \zeta} \frac{\partial B_{s}}{\partial \theta} \rb\\
    &-\frac{\mu}{B} \lp B^{\theta} \frac{\partial B}{\partial \theta} + B^{\zeta} \frac{\partial B}{\partial \zeta} \rp
\end{align}
in the domain $U \times \R$ where $U \subset \R^{3}$ is the stellarator volume. In flux coordinates, $U$ is represented by $[0,1] \times [0,2\pi] \times [0,2\pi]$. The terms $(B_{s}, B_{\theta}, B_{\zeta})$ and $(B^{\theta}, B^{\zeta})$ are the covariant and contravariant components of the magnetic field $\vb{B}(\vb{r})$, $B = |B(\vb{r})|$, and $\sqrt{g} = \det \lp \frac{\partial \vb{r}}{\partial(s,\theta,\zeta)} \rp$ is the Jacobian of the transformation from flux coordinates to Cartesian coordinates. Additionally, $\mu$ is the magnetic moment of the particle and $\Omega=2eB_{0}/m$ is a characteristic gyrofrequency for the alpha particles within the stellarator. Here $B_{0}$ represents a characteristic field strength, which we take to be the field strength at $s=0$, $\theta=0$, $\zeta=0$. Given initial conditions, we solve equations (\ref{guiding center}) up to $T_{\text{final}}$, and consider a particle confined if $s(t) < 1$ for $t < T_{\text{final}}$ and lost otherwise. 

\subsection{High expense of simulating energetic particle dynamics} \label{section:high-expense}

Considering Eq.\eqref{guiding center}, one may at first wonder why Monte Carlo estimators need to be improved. Indeed, Eq.\eqref{guiding center} ``only" is a 4-dimensional system of coupled ODEs, and since the equations are not modified by the presence of the other energetic particles considered in the Monte Carlo analysis, the calculations are embarrassingly parallelizable. It would be natural to conclude that the slow convergence of Monte Carlo estimators can be easily offset by the relative ease of brute-force computations. We briefly review here why this is not the case, and why direct computations of energetic particle losses remain too expensive for many design optimization applications. The reasons are intrinsically tied to the nature of the motion of energetic particles in non-axisymmetric stellarator magnetic fields, as well as to the numerical cost of evaluating the terms on the right-hand side of Eq.\eqref{guiding center}. 

\begin{figure}
    \centering
    \includegraphics*[trim = 250 20 200 50, clip, scale=0.34]{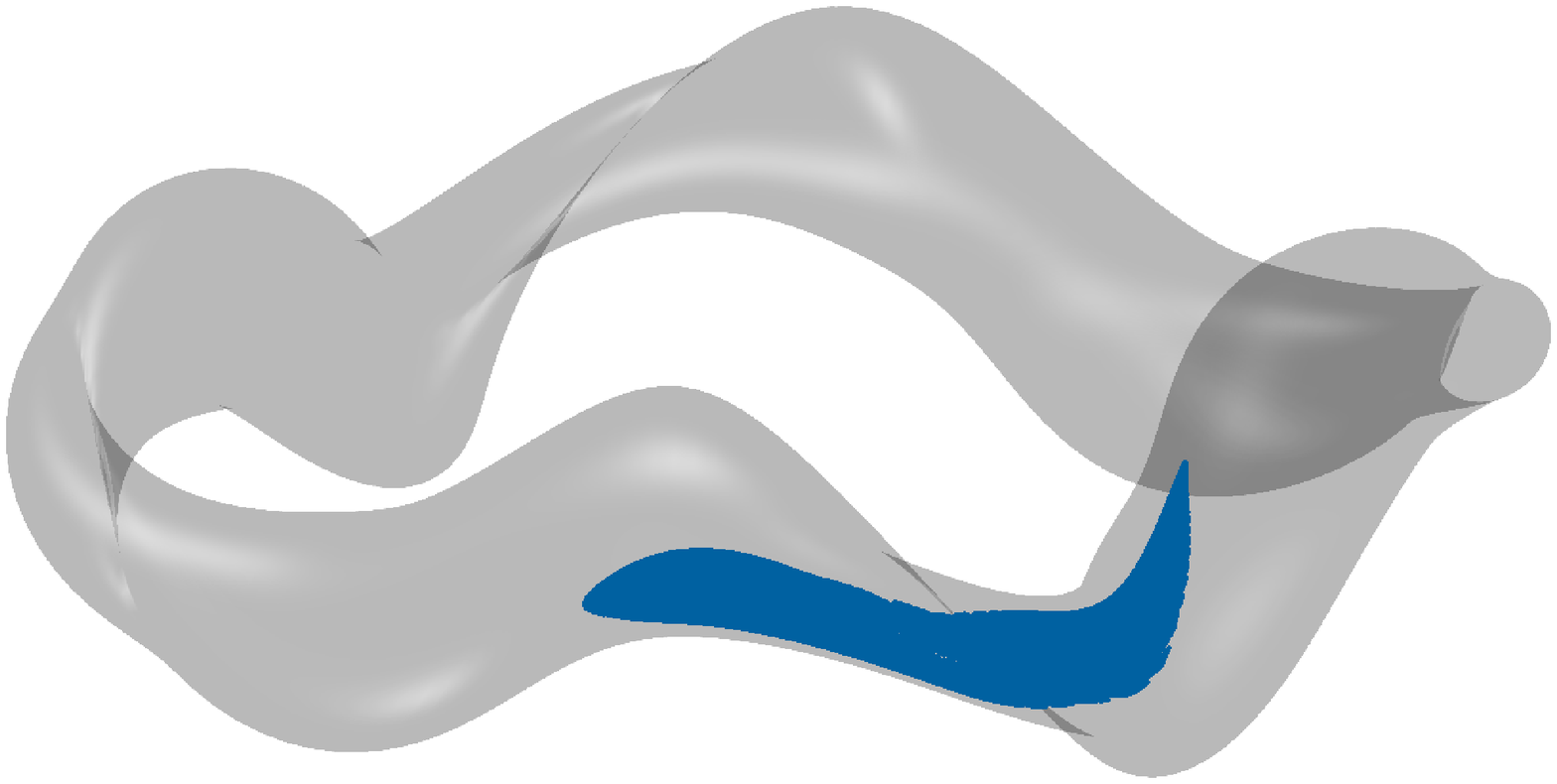} \includegraphics[scale=0.28]{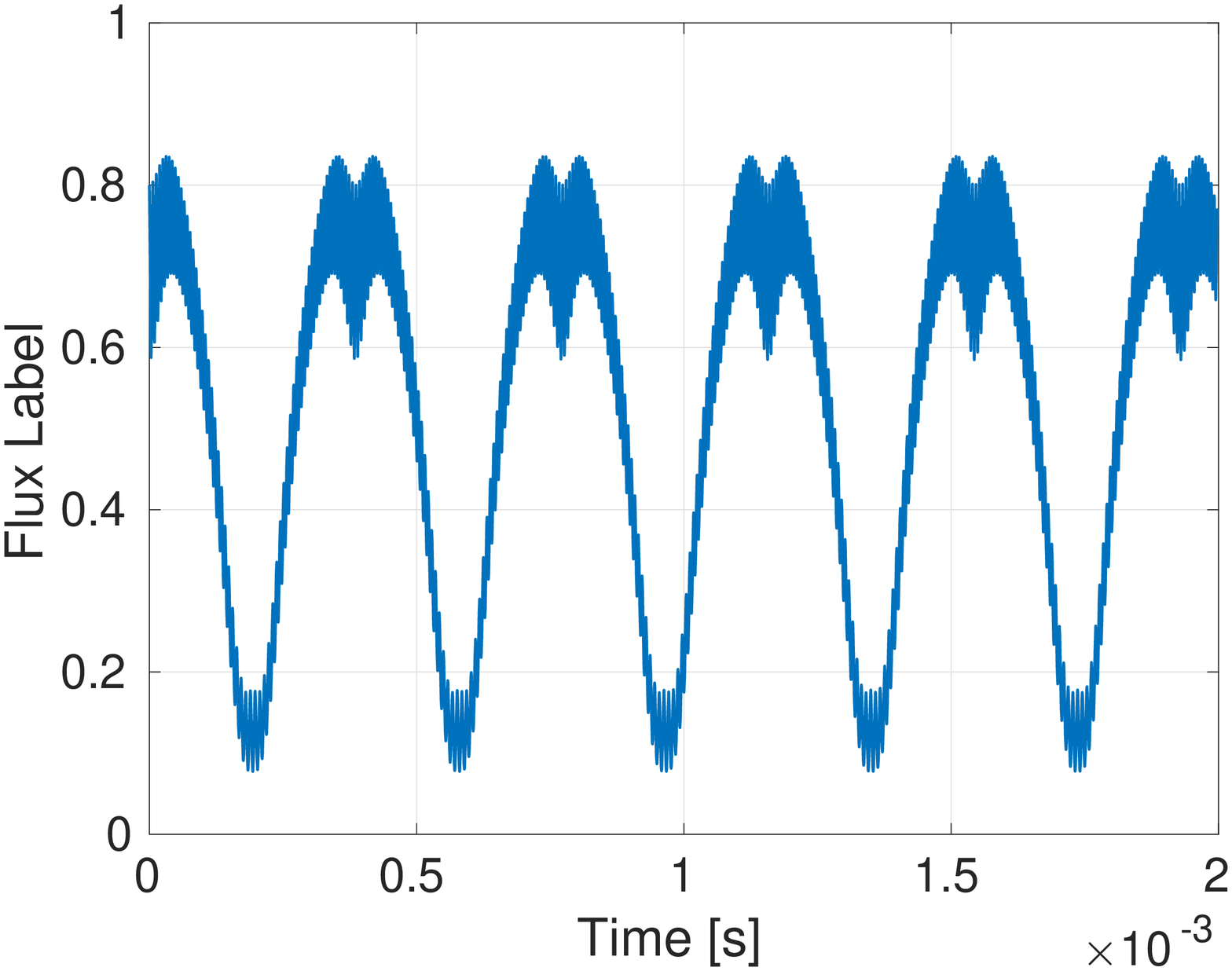}\\
    \includegraphics*[trim = 250 20 200 50, clip, scale=0.34]{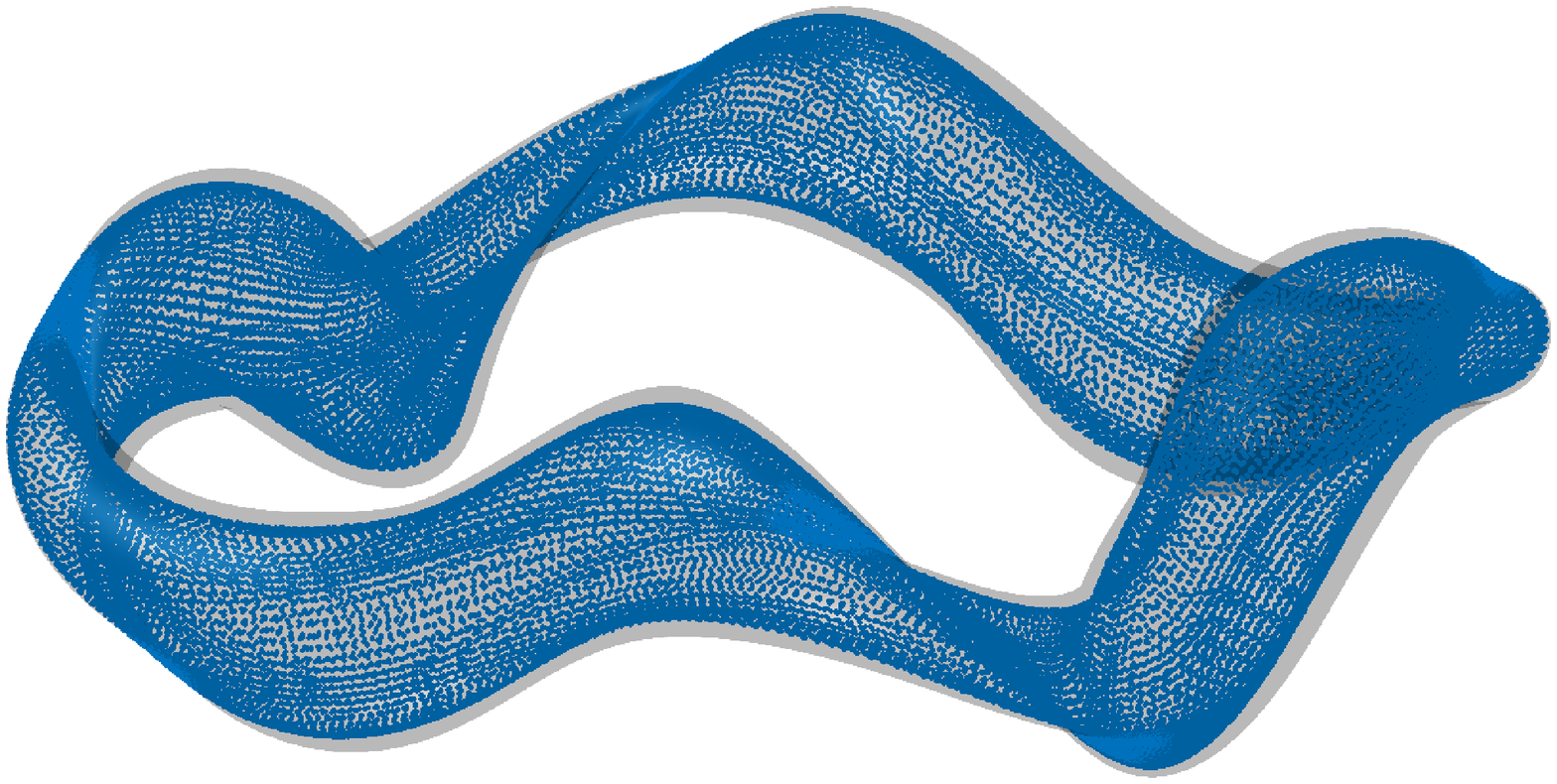} \includegraphics[scale=0.28]{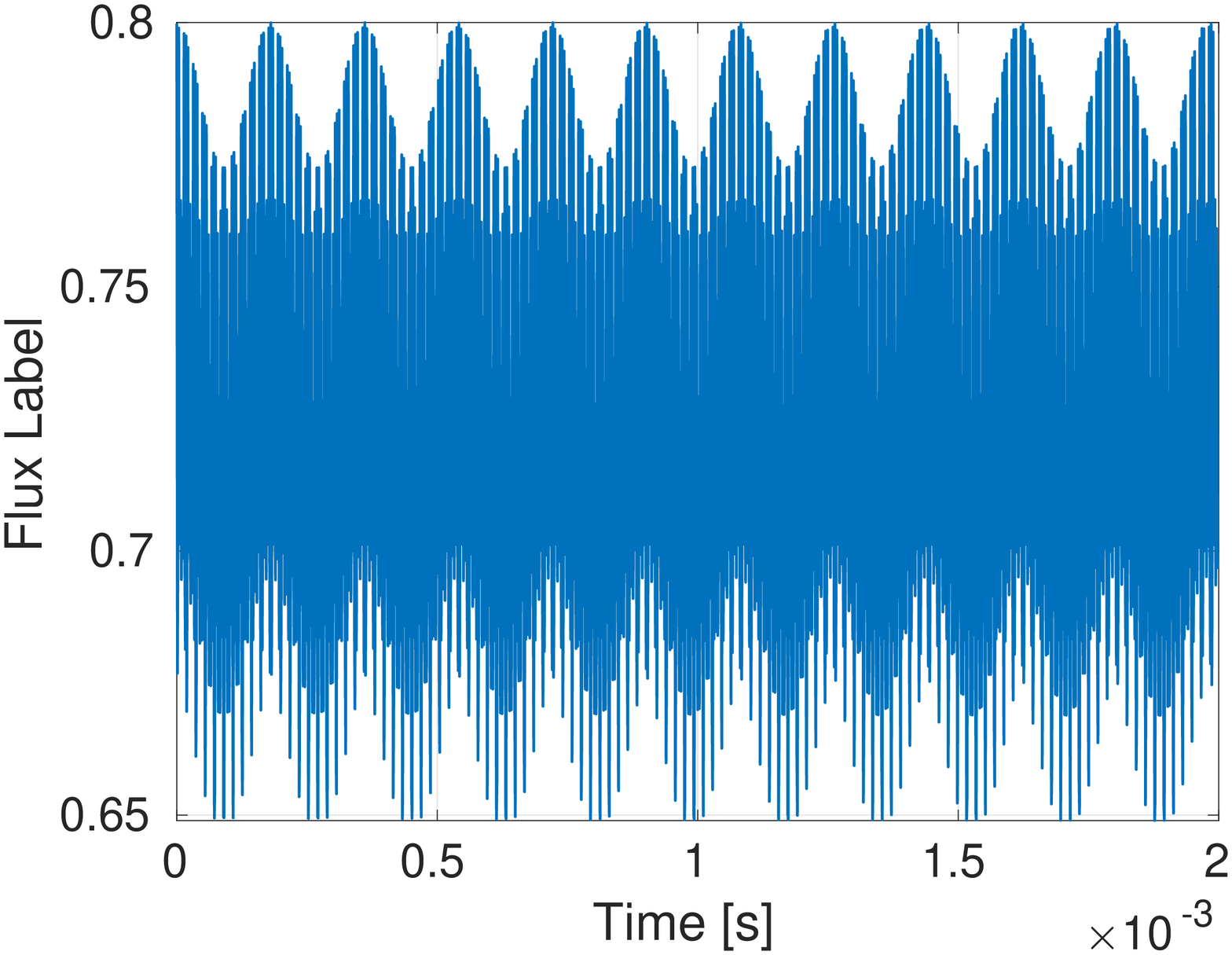}
    \caption{3D trajectories and flux label vs time for trapped (top) and passing (bottom) particles launched from the same location. The flux label for trapped and passing particles exhibit significantly different temporal and spatial scales.}
    \label{fig:time-scale-difference}
\end{figure}

As we discussed previously, energetic particle confinement analyses require the simulation of their dynamics for a time at least as long as their thermalization time \cite{Ku2008,Bader:2019}, after which their orbits become comparable to those of the background thermal ions. This energetic particle thermalization time, corresponding to tens of thousands of toroidal transits for passing particles \cite{Mynick2006}, and thousands of bounces for trapped particles, is significantly longer than the characteristic time scales of their single-particle dynamics. Because of the fast cross-field drift of the energetic particles and the high sensitivity of the orbits to local variations of the magnetic field, as illustrated in Figure \ref{fig:time-scale-difference}, the motion on these short time scales needs to be resolved with high accuracy. One is therefore faced with an intrinsically multi-scale problem, which is made even more difficult by the chaotic nature of some of the particle trajectories \cite{Albert:2020}. Numerically, this means that short time steps must be chosen when integrating Eq.\eqref{guiding center} to accurately resolve the dynamics on the short time scales. At the same time, high order integration schemes must be implemented, in order to prevent the numerical error from accumulating on the long thermalization time scale. The design of efficient and inexpensive numerical schemes satisfying this stringent criteria remains an open problem in computational science. Progress has been made for certain classes of multi-scale problems \cite{Fatkullin2004,E2009,Lee2014}, but the potential of these methods to speed-up guiding center calculations for energetic particles has not been proven. To this day, integrating Eq.\eqref{guiding center} with high accuracy on the long themalization time scale for a single energetic particle remains an intrinsically expensive problem. 

Regardless of the choice of numerical ODE integrator, the cost of solving equations (\ref{guiding center}) numerically is further increased by practical considerations regarding the evaluation of the functions $B_{s}, B_{\theta}, B_{\zeta}, B^{\theta}, B^{\zeta}, B, \text{ and } \sqrt{g}$ appearing on the right-hand side of \eqref{guiding center} at each time step. For stellarators, these functions are often only accessible as numerical outputs of a 3D field solver (vacuum or finite pressure MHD solve) 
\cite{Hirshman:1983,Merkel:1987,Reiman:1986}. They are typically represented by a double Fourier representation in $(\theta,\zeta)$ with an interpolant or spline in $s$ for the coefficients, and are accessed in the form:
\begin{align}
    f(s,\theta,\zeta) = \sum_{m,n}  f_{mn}(s) \cos (m\theta -n \zeta) \quad \text{or} \quad f(s,\theta,\zeta) = \sum_{m,n}  f_{mn}(s) \sin (m\theta -n \zeta). \label{eqn:doubleFourier}
\end{align}
As a result, evaluating the right-hand side of equations (\ref{guiding center}) may take a nontrivial amount of time. 

\section{Estimating energetic particle losses with standard Monte Carlo} \label{section:UQ-setup}
In this section, we first explain how we mathematically treat the uncertainty concerning the initial conditions required to compute the single particle motion according to Eq.\eqref{guiding center}, corresponding to the uncertainty in the location and velocity of the energetic particles at birth. We then present three different quantities we consider in this article to characterize energetic particle confinement. The first quantity is the energetic particle loss fraction, which is the standard figure of merit in optimization studies. The second and third quantities are the mean loss time and the mean flux surface position, which we motivate physically below. Finally, we briefly review the direct approach based on the standard Monte Carlo method for estimating these quantities.

\subsection{Uncertain initial conditions for guiding center dynamics}
The energetic particles we consider are alpha particles, born as byproducts of deuterium-tritium fusion reactions. There is intrinsic uncertainty in where and in what direction the particles are emitted, as well as in their kinetic energy at birth. This manifests as uncertainty in the initial conditions needed to solve the guiding center equations (\ref{guiding center}). 

In this work, we follow the standard practice of assuming that all alpha particles are born with the same kinetic energy of $3.5$ MeV \cite{Henneberg_2019,Bader:2019, Albert:2020}. This is a simplification of the actual alpha particle birth process. For accurate alpha particle confinement results, one should in principle account for the energy spread in the alpha birth energy, due to the kinematics of the collision process \cite{Brysk_1973,Zweben_1992,Heidbrink_1994,Appelbe_2011}. As we make this simplifying assumption, only two sources of uncertainty remain: the location of birth of alpha particles, and the velocity direction at birth. We model this uncertainty by imposing probability distributions on the initial position $\vb{R}_{0}$ and initial parallel velocity $V_{0}$. Specifically, we model alpha particles as being born uniformly in the stellarator volume $U$:
\begin{align}
    \vb{R}_{0} \sim \text{Uniform}(U). \label{input-uncertainty1}
\end{align}
We choose this model for its simplicity. It is physically relevant for flat density and temperature profiles, with a steep pedestal at the edge. For standard density and temperature profiles peaked on axis, our model overestimates the production of alpha particles at large radii. Consequently, the fraction of lost alpha particles in our setup is likely higher than those sought in practice.
The method based on MFMC that we present below can be adapted to any choice of probability distribution for $\vb{R}_{0}$, and we will discuss later in this article the extent to which the numerical results we present could be affected by the choice of distribution. For the emission direction, we use the convention that particles are born isotropically in velocity space \cite{Bader:2019,Albert:2020}:
\begin{align}
   V_{0} \sim \text{Uniform}(-V_{\text{max}}, V_{\text{max}}). \label{input-uncertainty2}
\end{align}
where $V_{\text{max}}$ is the velocity of a particle with a kinetic energy of $3.5$ MeV. 

\subsection{Metrics of energetic particle confinement}

In order to estimate energetic particle losses, we must propagate the initial uncertainty through the guiding center dynamics to classify which initial conditions lead to lost particles. Particle loss is thus measured by $\E \lb F^{\text{lost}} \lp \vb{R}_{0}, V_{0} \rp \rb$ where
\begin{align}
    F^{\text{lost}}(\vb{R}_{0}, V_{0}) := \begin{cases} 1, \quad \inf \{t \, : \, s(t; \vb{R}_{0}, V_{0}) = 1 \} < T_\text{final}\\
    0, \quad \text{otherwise} \label{classify-fn}
    \end{cases}
\end{align}
and $s(t; \vb{R}_{0}, V_{0})$ comes from solving equations (\ref{guiding center}) using the initial conditions $(\vb{R}_{0}, V_{0})$. Details on how $\vb{R}_{0}$ is converted to flux coordinates are discussed later in Section \ref{section:numerics}. Since the dynamics are deterministic, we can interpret $F^{\text{lost}}(\vb{R}_{0}, V_{0})$ as a Bernoulli random variable which classifies whether or not a particle is lost by time $T_{\text{final}}$ based on where and in what direction it is born.

$F^{\text{lost}}$ often is the key figure of merit considered in alpha particle confinement studies \cite{Lotz_1992,Mynick2006,Ku_2006,Najmabadi_2008,Henneberg_2019,Bader:2019,Albert:2020}. It corresponds to a discrete random variable. To demonstrate the applicability of our method to figures of merit corresponding to continuous random variables, we consider two other metrics which may be used for design optimization, and provide insights on a stellarator's efficacy in confining energetic particles. The first quantity of interest is the average time of loss of energetic particles. Its value for design optimization and performance analysis can be explained as follows. First, prompt losses, corresponding to alpha particles lost after a few transits or bounces, before they have significantly slowed down, lead to more damage to the first wall than slower losses. Second, alpha particles need to be confined for long enough that they heat the plasma as they slow down. Given two magnetic configurations with the same loss fraction after an alpha particle slowing down time, the more desirable one is the one with the longer mean time of loss. This figure of merit is also relevant for assessing the validity of the assumptions and models used for the analysis. If the order of magnitude of the mean time of loss is comparable or only slightly smaller than the alpha particle slowing down time, a large number of alpha particles remain confined long enough to be thermalized. Results obtained by simulating mono-energetic collisionless guiding center trajectories, as often done \cite{Henneberg_2019, Bader:2019,Albert:2020} and as we do for illustrative purposes in this work, should then be confirmed with codes including more detailed physics, and collisions in particular \cite{Drevlak_2014,Pfefferle_2016}. As we consider this figure of merit, we however have to deal with a mathematical difficulty, associated with the fact that for fully confined alpha particles, the time of loss is infinity. We circumvent this difficulty by defining a modified loss time, such that all particles are considered "lost" by $T_{\text{final}}$:
\begin{align}
    F^{\text{time}}(\vb{R}_{0}, V_{0}) := \min \lp \inf \{t \, : \, s(t; \vb{R}_{0}, V_{0}) = 1 \}, T_{\text{final}} \rp \label{losstime-fn}
\end{align}

Our second physical quantity of interest associated with a continuous random variable is the mean flux surface location of energetic particles. This quantity is not directly related to alpha particle loss, but is a complementary measure of the quality of energetic particle confinement. It provides physical insights on why certain configurations have lower losses, as well as information on the location of alpha power deposition. We will thus consider the quantity:
\begin{align}
    F^{\text{flux}}(t,\vb{R}_{0}, V_{0}) := s(t; \vb{R}_{0}, V_{0}) \label{flux-fn}
\end{align}
Therefore the \textit{mean modified loss time} $\E \lb F^{\text{time}} \lp \vb{R}_{0}, V_{0} \rp \rb$ and \textit{mean normalized flux surface label} $\E \lb F^{\text{flux}} \lp t,\vb{R}_{0}, V_{0} \rp \rb$ serve as alternative statistical metrics for a stellarator's effectiveness in confining energetic particles in our study. For the remainder of this article, we will take $F$ to represent either $F^{\text{lost}}$, $F^{\text{time}}$, or $F^{\text{flux}}$ at some fixed time $t$.

\subsection{Uncertainty propagation using Monte Carlo}
Now that we have identified energetic particle loss, along with two other metrics for energetic particle confinement, as an expectation of a function $F(\vb{R}_{0}, V_{0})$, it remains to accurately estimate that expectation. A direct method for this is the Monte Carlo (MC) estimator, whereby we draw $p$ i.i.d samples $\{ ( \vb{R}_{0}^{(i)}, V_{0}^{(i)} )\}_{i=1}^{p}$ from (\ref{input-uncertainty1})-(\ref{input-uncertainty2}) and then approximate $\E[F]$ by the unbiased estimator
\begin{align}
    \wh{F}_{\text{MC},p} := \frac{1}{p} \sum_{i=1}^{p} F(\vb{R}_{0}^{(i)},V_{0}^{(i)}). \label{MC-estimator}
\end{align}
This standard Monte Carlo approach is summarized in Method \ref{MC-summary}. Since the MC estimator has variance $\Var(F)/p$, it converges at the slow rate of $1/\sqrt{p}$ with respect to the root mean square error. In order to compute the MC estimator, we need to solve the guiding center equations (\ref{guiding center}) a total of $p$ times and then evaluate $F$ using those trajectories. Due to the high computational cost of simulating guiding center trajectories of energetic particles over the appropriate time scale, as explained in Section \ref{section:high-expense}, using the MC estimator can rapidly become computationally intractable for large $p$.

The large cost required to produce an accurate MC estimator motivates us to aim for an estimator with lower variance via variance reduction. Such a procedure would lower the constant in front of $1/\sqrt{p}$, thereby yielding a more accurate estimator for the same computational effort.
\\
\begin{pabox}[label={MC-summary},nameref={}]{Monte Carlo estimator}
\begin{enumerate}
	\item Draw $p$ i.i.d. samples $(\vb{R}_{0}^{(1)}, V_{0}^{(1)}), (\vb{R}_{0}^{(2)}, V_{0}^{(2)}), \dotsc, (\vb{R}_{0}^{(p)}, V_{0}^{(p)})$ from (\ref{input-uncertainty1})-(\ref{input-uncertainty2}).
	\item For each $i=1,\dotsc,p$, solve the guiding center equations (\ref{guiding center}) using expensive numerical integration. Use the computed trajectories to evaluate $F(\vb{R}_{0}^{(i)},V_{0}^{(i)})$ for $i=1,\dotsc,p$.
	\item Estimate $\E[F]$ by
	\begin{align*}
	    \wh{F}_{\text{MC},p} := \frac{1}{p} \sum_{i=1}^{p} F(\vb{R}_{0}^{(i)},V_{0}^{(i)}). 
	\end{align*}
\end{enumerate}
\end{pabox}

\section{Estimating energetic particle confinement with multifidelity Monte Carlo} \label{section:MFMC}
Here, we first describe how one can leverage a surrogate model to construct a control variate based multifidelity estimator with smaller variance than the standard Monte Carlo estimator. We then discuss an approach to construct highly correlated surrogate models for the analysis of energetic particle confinement.

\subsection{Variance reduction with multifidelity Monte Carlo}

Alongside the high-fidelity model $F$, which requires expensive numerical integration to evaluate, suppose we also have a low-fidelity, or \textit{surrogate}, model $G$. In order to leverage $G$ for variance reduction we utilize a control variate based MFMC method \cite{Peherstorfer:2016}. The MFMC estimator with computational budget $p$ is the unbiased estimator
\begin{align}
        \wh{F}_{\text{MFMC},p} = \wh{F}_{\text{MC},n} + \alpha \lp \wh{G}_{\text{MC},m} - \wh{G}_{\text{MC},n} \rp \label{MFMC-est}
\end{align}
where $\alpha$ is a constant, $n$ is the number of high-fidelity samples and $m$ is the number of low-fidelity samples. To minimize the variance of the MFMC estimator, $\alpha,n,m$ can be optimally chosen, and those choices satisfy
\begin{align}
    \alpha = \rho(F,G) \sqrt{\frac{\Var(F)}{\Var(G)}}, \quad p = n + \frac{m}{w}, \quad m = n \sqrt{\frac{w\rho(F,G)^{2}}{1-\rho(F,G)^{2}}} \label{optimal-MFMC-params}
\end{align}
where $\rho(F,G)$ is the Pearson's correlation coefficient between the high-fidelity $F$ and the surrogate $G$ and $w$ is the ratio of cost of evaluating $F$ to the cost of evaluating $G$. A practical implementation of this method is summarized in Method \ref{MFMC-summary}. The variance of the resulting optimal MFMC estimator is given by
\begin{align}
    \Var \lp \wh{F}_{\text{MFMC},p} \rp = \lb \lp \sqrt{1 - \rho(F,G)^{2}} + \sqrt{\frac{\rho(F,G)^{2}}{w}} \rp^{2} \rb \Var \lp \wh{F}_{\text{MC},p} \rp.  \label{MFMC-variance}
\end{align} 
In order to maximize the amount of variance reduction MFMC provides compared to MC, the bracketed term in (\ref{MFMC-variance}) should be as small as possible. This means that we want the surrogate $G$ to be both highly correlated to $F$, as well as much cheaper to evaluate than $F$. Although the MFMC estimator still converges at the same slow rate of $1/\sqrt{p}$ as the standard MC estimator, it can be more accurate with the same computational budget provided a sufficiently correlated and inexpensive surrogate.

Note that once one is given a surrogate $G$, the amount of speedup provided by MFMC is constant. That is, the construction of $G$ is done offline, separate from the online phase of estimation. No matter how expensive the construction of $G$ is (which could require many samples of $F$ for instance), once we have that surrogate, the speedup from MFMC is guaranteed no matter how large $p$ is taken to be; see \cite{P19AMFMC} for an extension of MFMC that takes offline costs into account

In the MFMC literature, the cost of sampling the initial uncertainty is not taken into account, as it is typically assumed to be orders of magnitude cheaper than everything else being considered. In practice, if sampling the input uncertainty requires non-trivial cost, then a better definition of the cost ratio $w$ would be required as the current definition only accounts for the cost of propagating the uncertainty forward. While we do not account for this cost of input uncertainty to make a fair budget comparison in our MFMC experiments, we mention here the limits on implementation imposed by how the input uncertainty is sampled.
\begin{pabox}[label={MFMC-summary},nameref={}]{Multifidelity Monte Carlo estimator}
\begin{enumerate}
	\item Given $p$, compute $n$ and $m$ as in (\ref{optimal-MFMC-params})
    \footnote{Use an estimate $\hat{\rho}$ of $\rho(F,G)$. If an estimate is not readily available, one can be found easily using samples}. 
	Draw $m$ i.i.d. samples $(\vb{R}_{0}^{(1)}, V_{0}^{(1)}), (\vb{R}_{0}^{(2)}, V_{0}^{(2)}), \dotsc, (\vb{R}_{0}^{(m)}, V_{0}^{(m)})$ from (\ref{input-uncertainty1})-(\ref{input-uncertainty2}). 
	\item For each $i=1,\dotsc,n$, solve the guiding center equations (\ref{guiding center}) using expensive numerical integration. Use the computed trajectories to evaluate $F(\vb{R}_{0}^{(i)},V_{0}^{(i)})$ for $i=1,\dotsc,n$. Evaluate $G(\vb{R}_{0}^{(i)},V_{0}^{(i)})$ for $i=1,\dotsc,m$.
	\item Set $\alpha$ as in (\ref{optimal-MFMC-params}), replacing $\Var(F)$ and $\Var(G)$ with sample variances computed using the evaluations from (ii).
	\item Estimate $\E[F]$ by
	\begin{align}
	    \wh{F}_{\text{MFMC},p} = \wh{F}_{\text{MC},n} + \alpha \lp \wh{G}_{\text{MC},m} - \wh{G}_{\text{MC},n} \rp.
	\end{align}
\end{enumerate}
\end{pabox}

\subsection{Choice of surrogate model} \label{section:surrogate-choice}

We now detail our choice of a data-fit surrogate model $G$ designed to maximize correlation with $F$ for our estimation of energetic particle confinement. We rely on data-fit surrogates because traditional sources of surrogates are not appropriate for energetic particle dynamics. For example, there is no natural hierarchy of simplified physics models to leverage \cite{Peherstorfer:2018}, since the guiding center equations are already a simplification of the computationally intractable full-orbit dynamics, and a "beads on the wire"  model for the particle trajectories, which would omit all finite-gyroradius corrections, as in the kinetic MHD model \cite{KruskalOberman,RosenbluthRostokerKineticMHD,GradGuidingCenter,waelbroeck2018framework,CerfonFreidbergKineticMHD,Ramos1,Ramos2,BurbySenguptaKineticMHD}, would not lead to good correlation since it neglects the large cross-field drifts of energetic particles. Similarly, we have empirically found that another usually powerful source of surrogates, arising from successively coarsening the grid \cite{Cliffe:2011}, is not satisfactory either for our application. That is because the time step of the numerical ODE integrator is tightly constrained by the requirement to resolve the shortest characteristic time scales of the motion of both trapped and passing particles. When this time step is significantly increased, accuracy is significantly reduced, and correlation rapidly decreases. Lastly, since the problem is transport dominated, popular surrogates from traditional model reduction, such as proper orthogonal decomposition or reduced basis methods, are a poor choice \cite{OhlbergerSuccess}.

For a data-fit surrogate, we consider a linear basis function model of the form:
\begin{align}
    G^{n}(\vb{R}_{0}, V_{0};\mathbf{P}^{n}) = \sum_{k=0}^{n} P^{n}_{k} \phi_{k}(\vb{R}_{0}, V_{0}) \label{lbf-models}
\end{align} 
where $\mathbf{P}^{n} \in \R^{n+1}$ and $\{\phi_{k}\}_{k=0}^{n}$ is any set of basis functions. Using this family of surrogates, we want to find the set of parameters $\mathbf{P}^{n}$ which maximizes the correlation with our high-fidelity model. This amounts to solving
\begin{align}
\underset{\mathbf{P}^{n}}{\text{maximize}}  \quad  \rho(F,G^{n}(\cdot \, ; \mathbf{P}^{n})) \label{C-opt}
\end{align}
However, a caveat of this formulation is that the optimization problem (\ref{C-opt}) has infinitely many maxima through scaling, since our model is linear, $ \rho(F,G^{n}(\cdot \, ; c \mathbf{P}^{n})) = \rho(F,c  G^{n}(\cdot \, ; \mathbf{P}^{n})) = \rho(F,G^{n}(\cdot \, ; \mathbf{P}^{n}))$ for any $c > 0$.

It thus is better to consider $\mathbf{P}^{n}_{\ast}$, which solves the following optimization problem (\ref{M-opt}):
\begin{align} 
\underset{\mathbf{P}^{n}}{\text{minimize}} \quad \mathbb{E}[(F-G^{n}(\cdot \, ; \mathbf{P}^{n}))^{2}] \label{M-opt}
\end{align} Due to the linear nature of our family of surrogate models, the unique solution $\mathbf{P}_{\ast}^{n}$ of (\ref{M-opt}) is also one of the non-unique solution for (\ref{C-opt}).
 
In practice, evaluating the objective function in \eqref{M-opt} is as challenging as the original problem motivating our work, since it involves evaluating expectations of $F$. Finding the exact minimum $\mathbf{P}_{\ast}^{n}$ is therefore a computationally intractable problem, regardless of the choice of basis functions $\{\phi_{k}\}_{k=0}^{n}$. We now explain how we address this difficulty and construct a satisfactory approximation of $\mathbf{P}_{\ast}^{n}$ for the particular choice of basis functions $\{\phi_{k}\}_{k=0}^{n}$ we make for this study, namely piece-wise linear functions. The idea is to temporarily change the interpretation of $F$, and view it as a deterministic map from the four-dimensional space $(s,\theta,\zeta,v_{||})$ to $\mathbb{R}$. First, in $(s,\theta,\zeta,v_{||})$ coordinates, we construct a regular mesh on $[0,1] \times [0,2\pi] \times [0,\pi/2] \times [-V_{\text{max}}, V_{\text{max}}]$, and evaluate $F$ on that mesh. Then we specifically choose our basis functions $\{\phi_{k}\}$ to be the hat functions on the mesh, so that our family of linear basis functions $G^{n}$ is the family of piece-wise linear functions on the mesh. If $F$ is a continuous map, and we determine $\wt{\mathbf{P}}^{n}$ so that $G^{n} (\cdot, \wt{\mathbf{P}}^{n})$ is the corresponding interpolant of $F$ on the mesh, then $\wt{\mathbf{P}}^{n}$ will be a close approximation to $\mathbf{P}_{\ast}^{n}$. The reasoning behind this is that if $F$ is continuous, as we refine the mesh then $G^{n}(\cdot \,; \mathbf{P}^{n}_{\ast})$, where $\mathbf{P}^{n}_{\ast}$ solves (\ref{M-opt}) on successively finer meshes, will converge to $F$. Since $G^{n}(\cdot\, ; \wt{\mathbf{{P}}}^{n})$ also converges to $F$ as the mesh is refined, we expect $\wt{\mathbf{{P}}}^{n}$ to be a strong approximation to $\mathbf{P}^{n}_{\ast}$, converging as the mesh refines. Although we cannot measure how close $\wt{\mathbf{P}}^{n}$ is to $\mathbf{P}^{n}_{\ast}$, in practice we have observed that $G^{n} (\cdot, \wt{\mathbf{P}}^{n})$ is favorably correlated with $F$, even on a relatively coarse mesh.

\section{Numerical results} \label{section:numerics}

We now conduct numerical tests to determine the level of variance reduction and the speed up one can obtain in practice with our MFMC approach in a realistic stellarator configuration. We first provide details for the setup of our numerical experiments. Then we demonstrate the effectiveness of MFMC for confinement studies done at high resolution, with a time step equal to the characteristic gyroperiod of alpha particles in our magnetic equilibrium. Since the time step is small, we cannot track particle orbits beyond $T_{\text{final}}=0.1\;\mbox{ms}$ with the computing resources available to us. This is well below the typical alpha particle slowing down time, which is approximately 200 $\;\mbox{ms}$. In terms of confinement, the results of this analysis are mostly dominated by prompt particle losses. In the final part of this section, we decrease the time resolution of our high-fidelity numerical integrator, in order to compute particle trajectories up to $T_{\text{final}}=20\;\mbox{ms}$. While this remains below the typical alpha particle slowing down time, it is long enough to include more varied particle loss channels, and thus provide realistic estimates of the effectiveness of MFMC for alpha particle confinement studies.

\subsection{Numerical Setup} \label{section:num-setup}
For our high-fidelity solver, we use a Runge-Kutta 4 (RK4) time integrator for the guiding center equations. The terms $B_{s}, B_{\theta}, B_{\zeta}, B^{\theta}, B^{\zeta}, \sqrt{g}$ are provided as numerical outputs of a VMEC solve \cite{Hirshman:1983}. These functions are given by the double Fourier representation (\ref{eqn:doubleFourier}) using 128 pairs $(m,n)$ in the $(\theta,\zeta)$ variables, where only cosine or sine series arise due to stellarator symmetry. For each $(m,n)$ pair, $f_{mn}(s)$ is represented using a smoothing spline over 101 equispaced points in $[0,1]$. We perform all tests on a Wistell-A vacuum magnetic configuration. Wistell-A is a quasihelically symmetric stellarator with 4-fold stellarator symmetry \cite{Bader2020new}. For the time step size in our RK4 integrator, we choose $\Delta \tau = 2\pi/\Omega$ which is the characteristic gyroperiod of alpha particles in the Wistell-A magnetic field. This step size allows us to accurately resolve the guiding center dynamics, but limits us in how large $T_{\text{final}}$ can be. 

We shall refer to our low-fidelity models as $G^{\text{lost}}$, $G^{\text{time}}$, and $G^{\text{flux}}$ for the quantities of interest $F^{\text{lost}}$, $F^{\text{time}}$, and $F^{\text{flux}}$, respectively. To track the flux label in time, we specify $N_{\text{flux}}$ time points $t_{k} = (\text{time step size})\times N_{\text{skip}} k$ for $k=1,\dotsc, N_{\text{flux}}$. Note that $N_{\text{flux}}$ and $N_{\text{skip}}$ change depending on the time step size, and are adjusted in order to get to the desired $T_{\text{final}}$. The reasoning behind recording the flux label every $N_{\text{skip}}$ steps is that for any practical value of $T_{\text{final}}$, tracking the trajectory at every step requires an unreasonably large number of specified time points. For different time step sizes, we choose $N_{\text{skip}}$ in an attempt to make the specified time points similar to one another. For example, if the time step size is halved, then $N_{\text{skip}}$ would be doubled. If the time step size is divided by three, then $N_{\text{skip}}$ would be tripled, and $N_{\text{flux}}$ adjusted as well.
Thus, for an input $(\vb{R}_{0}, V_{0})$, our high-fidelity flux label model outputs $\{F^{\text{flux}}(t_{k},\vb{R}_{0},V_{0})\}_{k=1}^{N_{\text{flux}}}$. Our low-fidelity flux label model is then $\{G^{\text{flux}}_{k}(\vb{R}_{0},V_{0})\}_{k=1}^{N_{\text{flux}}}$ where $G^{\text{flux}}_{k}$ is the piece-wise linear interpolant to $F^{\text{flux}}(t_{k},\cdot)$ over a regular mesh in $(s,\theta,\zeta,v_{||})$ as detailed in $\ref{section:surrogate-choice}$. Similarly $G^{\text{time}}$ is the piece-wise linear interpolant to $F^{\text{time}}$ over the same mesh. For all interpolants we use 25 points in each dimension. For $G^{\text{lost}}$ we specify an optimal threshold using $G^{\text{time}}$. Namely, we set
\begin{align}
    G^{\text{lost}}(\vb{R}_{0}, V_{0}) := \begin{cases} 1, \quad G^{\text{time}}(\vb{R}_{0}, V_{0}) \leq T_{\text{threshold}}\\
    0, \quad \text{otherwise} \label{low-fid-classify}
    \end{cases}
\end{align}
where $T_{\text{threshold}}$ is chosen to minimize the expected misclassification $\E\lb \lp F^{\text{lost}} - G^{\text{lost}} \rp^{2} \rb$. To compute $T_{\text{threshold}}$, we use a sample average approximation of the expected misclassification using $10^{4}$ samples, and then choose $T_{\text{threshold}}$ to be the minimizer of that sample average approximation evaluated at $10^{4}$ equispaced points between $0$ and $T_{\text{final}}$. 

\begin{table}[b]
    \def\arraystretch{1.3}
    \centering
    \begin{tabular}{l l l l l l}
    \hline
    (time step size, $T_{\text{final}}$ [ms]) & ($\Delta \tau$, 0.1) & ($3\Delta \tau$, 20) & ($5\Delta \tau$, 20) & ($7\Delta \tau$,20) & ($10\Delta \tau$, 20)\\
    \hline
    high-fidelity cost [s] & 20 & 1333 & 800 & 571 & 400\\
    cost ratio, $G^{\text{lost}}$ and $G^{\text{time}}$ & 2e$+$06 & 1.33e$+$08 & 8e$+$07 & 5.71e$+$07 & 4e$+$07\\
    cost ratio, $G^{\text{flux}}$ & 6e$+$04 & 4e$+$06 & 3.6e$+$06 & 1.7e$+$06 & 1.2e$+$06\\
    $N_{\text{flux}}$ & 684 & 760 & 684 & 978 & 684\\
    $N_{\text{skip}}$ & 100 & 300 & 200 & 100 & 100\\
    \hline
    \end{tabular}
    \caption{Costs for the high-fidelity model in seconds, cost ratios, and $N_{\text{flux}}$ and $N_{\text{skip}}$ used for various time step sizes and integration lengths. We see that for $T_{\text{final}}=20\;\mbox{ms}$ our low-fidelity models are several millions of times faster than the high-fidelity evaluation. Only the values for ($\Delta \tau$, 0.1) were measured, all other columns result from scaling.
    }
    \label{tab:costs}

\vspace{10mm}

    \begin{tabular}{l l l l l l}
     \hline
    (time step size, $T_{\text{final}}$ [ms]) & ($\Delta \tau$, 0.1) & ($3\Delta \tau$, 20) & ($5\Delta \tau$, 20) & ($7\Delta \tau$,20) & ($10\Delta \tau$, 20)\\
    \hline
    Classification of lost particles & 0.8776 & 0.9076 & 0.8996 & 0.8762 & 0.8460 \\
    Modified loss time & 0.9444 & 0.9299 & 0.9426 & 0.9436 & 0.9247\\
    Normalized flux label (average) & 0.9846 & 0.9538 & 0.9517 & 0.9403 & 0.9111\\
    Normalized flux label (final time) & 0.9761 & 0.9486 & 0.9452 & 0.9270 & 0.8833 \\
    \hline
    \end{tabular}
    \caption{Correlations between the high-fidelity and low-fidelity models for each quantity of interest using different time step sizes and $T_{\text{final}}$. For the normalized flux label we report both the mean correlation over time points as well as the correlation at $T_{\text{final}}$. Correlations are measured using $10^{4}$ samples.}
    \label{tab:correlation}
\end{table}

To sample $\vb{R}_{0} \sim U$ in magnetic coordinates, we take advantage of the 4-fold stellarator symmetry in Wistell-A and sample $(s_{0},\theta_{0},\zeta_{0})$ from the probability density proportional to $\sqrt{g}$ over $D=(0,1) \times (0,2\pi) \times (0,\pi/2)$. This is done by numerically integrating $\sqrt{g}$ over $D$ to compute the normalization constant, and then performing rejection sampling with a uniform proposal. This step is relatively fast and requires on average only 3 rejections before a successful draw from the target distribution.

The costs of evaluating $F^{\text{lost}}$, $F^{\text{time}}$, and $F^{\text{flux}}$ are all the same: the time required to numerically integrate to $T_{\text{final}}$ using a given time step size. The costs of evaluating $G^{\text{lost}}$ and $G^{\text{time}}$ are the same, but the cost of evaluating $G^{\text{flux}}$ is slightly larger since it requires evaluating $N_{\text{flux}}$ interpolants. In Table \ref{tab:costs} we report the costs of evaluating the high-fidelity model, the cost ratios for $G^{\text{lost}}$, $G^{\text{time}}$, and the cost ratios for $G^{\text{flux}}$ across multiple time step sizes and values of $T_{\text{final}}$. We note that for $T_{\text{final}}=20\;\mbox{ms}$ the low-fidelity models are all several millions of times faster than the high-fidelity models. 

In Table \ref{tab:correlation} we report the correlations between our high and low-fidelity models for each of the quantities of interest across multiple time step sizes and values of $T_{\text{final}}$. We measure the correlation coefficient using $10^{4}$ samples.

\subsection{High resolution high-fidelity model and short time integration}

We numerically compare the MC and MFMC estimators for the loss fraction, mean modified loss time, and mean flux label using our low-fidelity models for a situation in which we have a high resolution high fidelity time integrator, with which we can only afford to compute trajectories over a relatively short time. Specifically, here we use time step size $\Delta \tau$ and choose $T_{\text{final}}$ as 0.1ms. For computational budgets $p=100,250,500,1000$, we measure the root mean square error (RMSE) of MC and MFMC by using 100 replicates of each estimator and compare them to a reference solution generated using MC with $10^{6}$ samples. 

\begin{figure}[!t]
    \centering
    \includegraphics[scale=0.32]{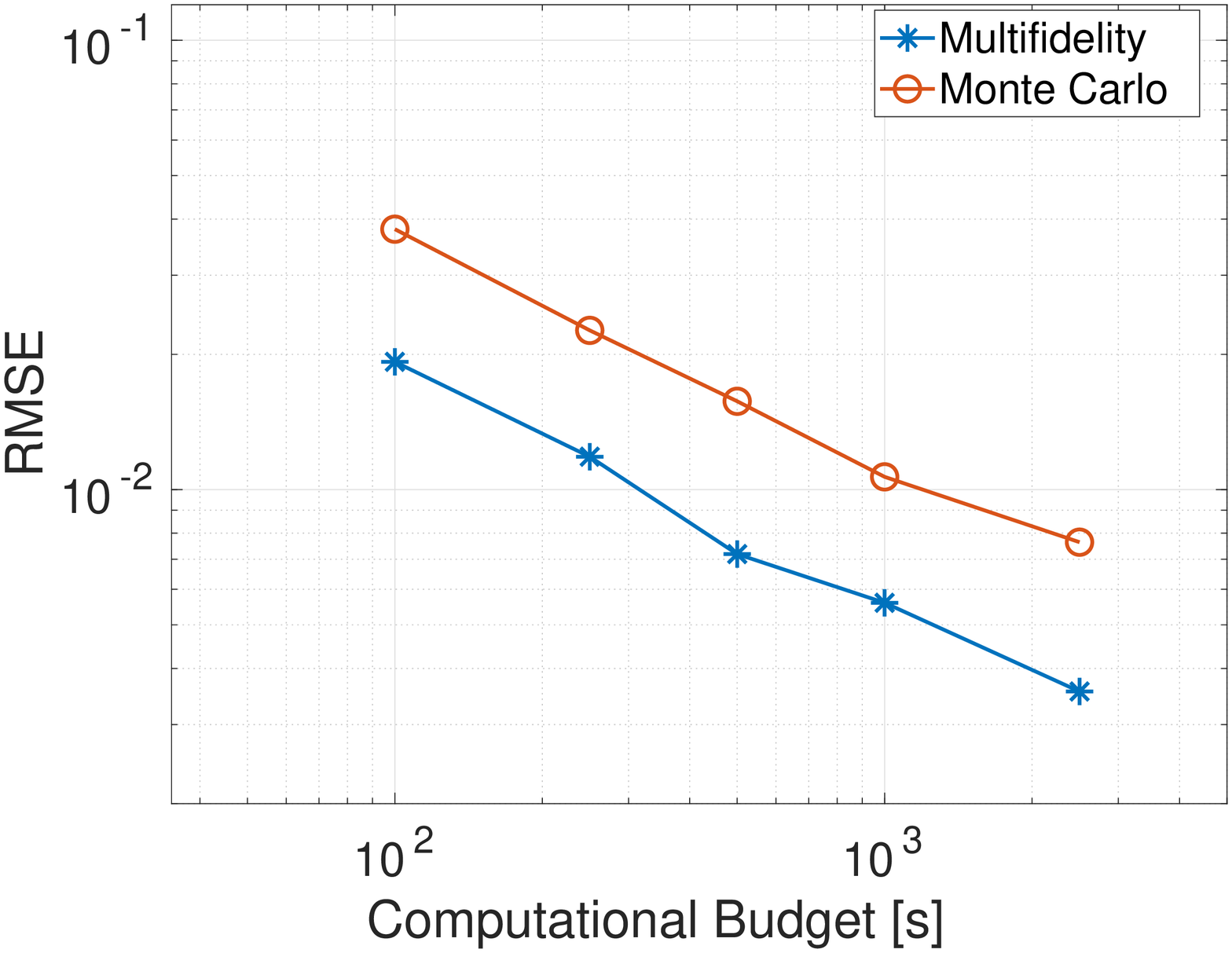}
    \includegraphics[scale=0.32]{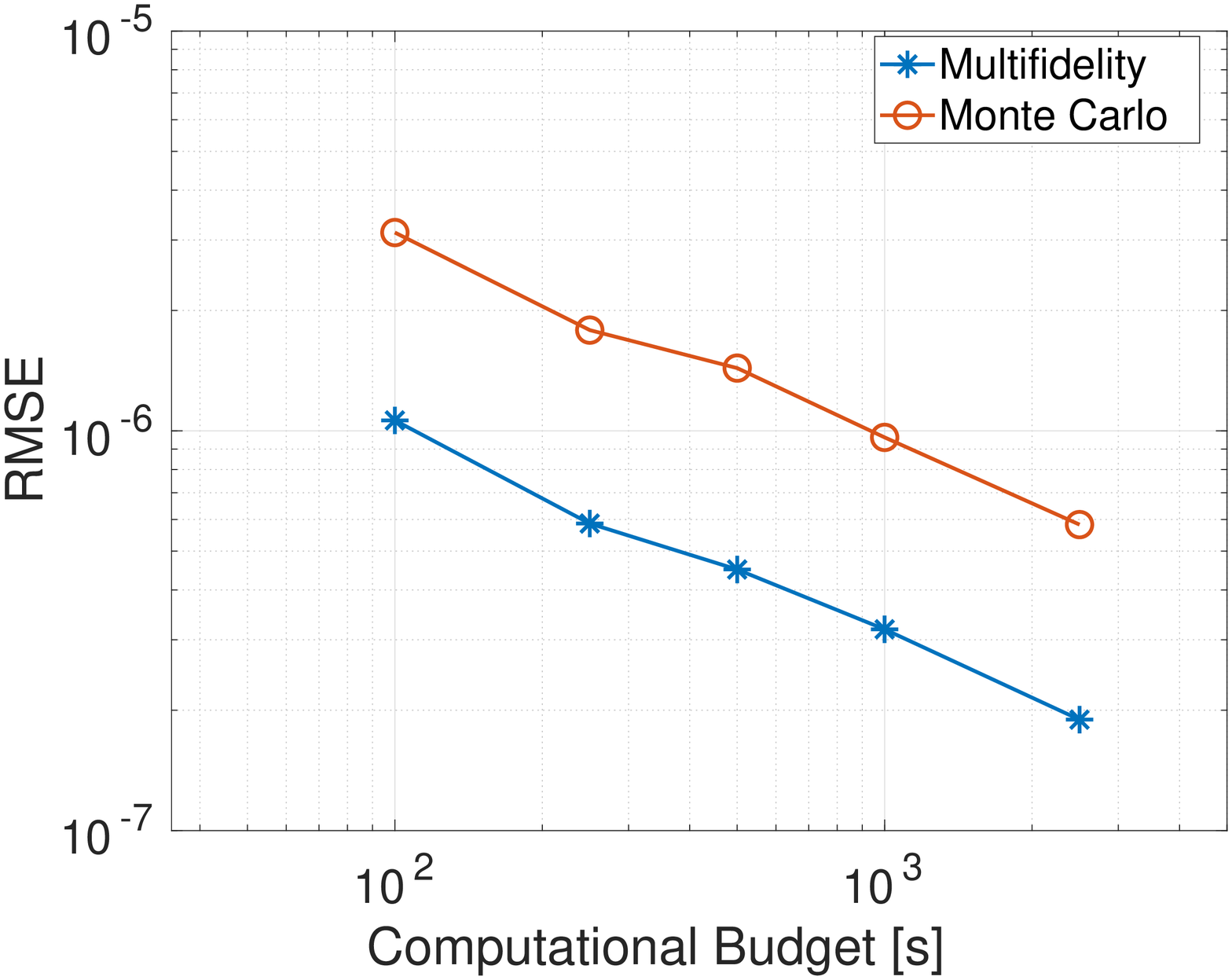}\\
    \includegraphics[scale=0.32]{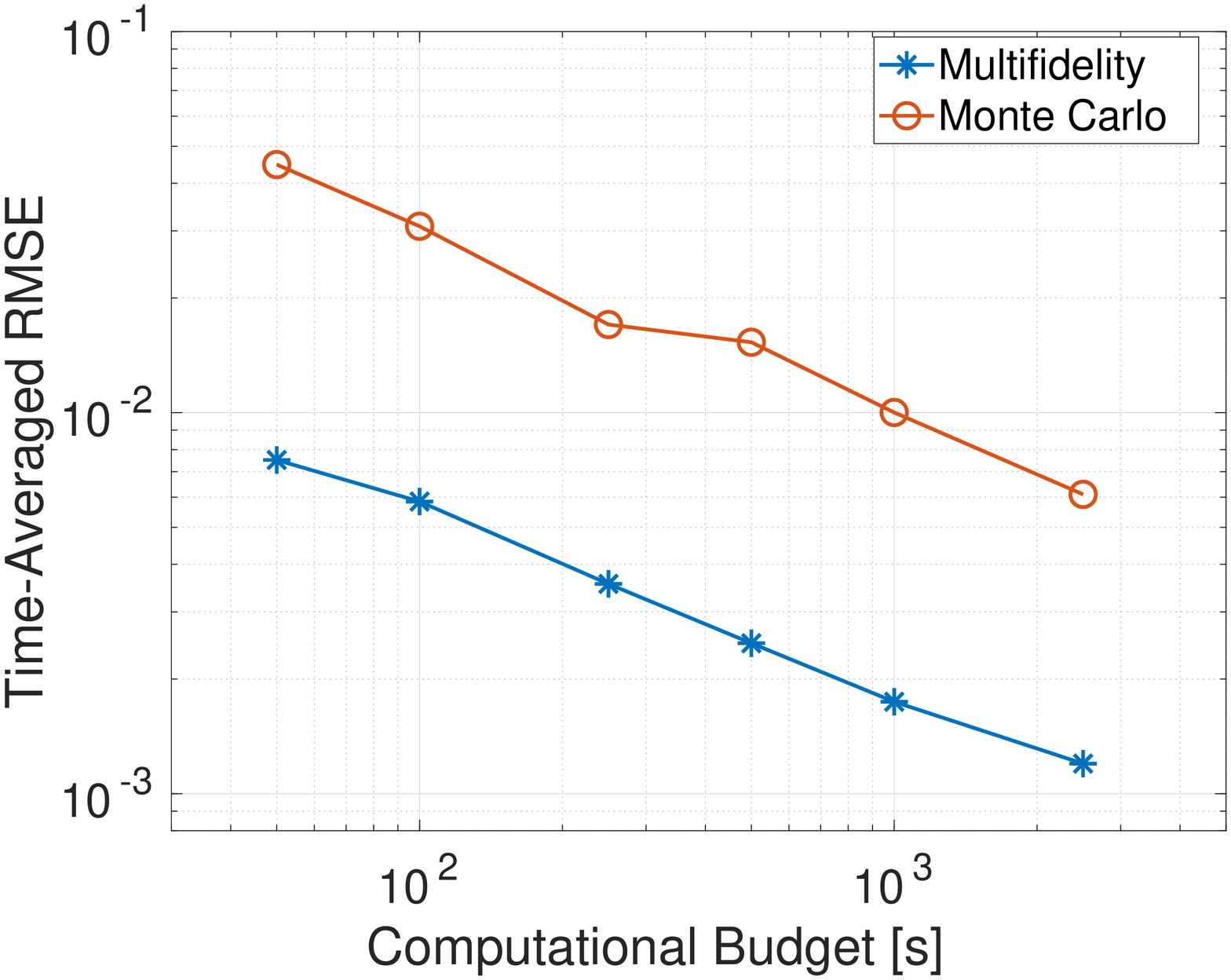}
    \includegraphics[scale=0.32]{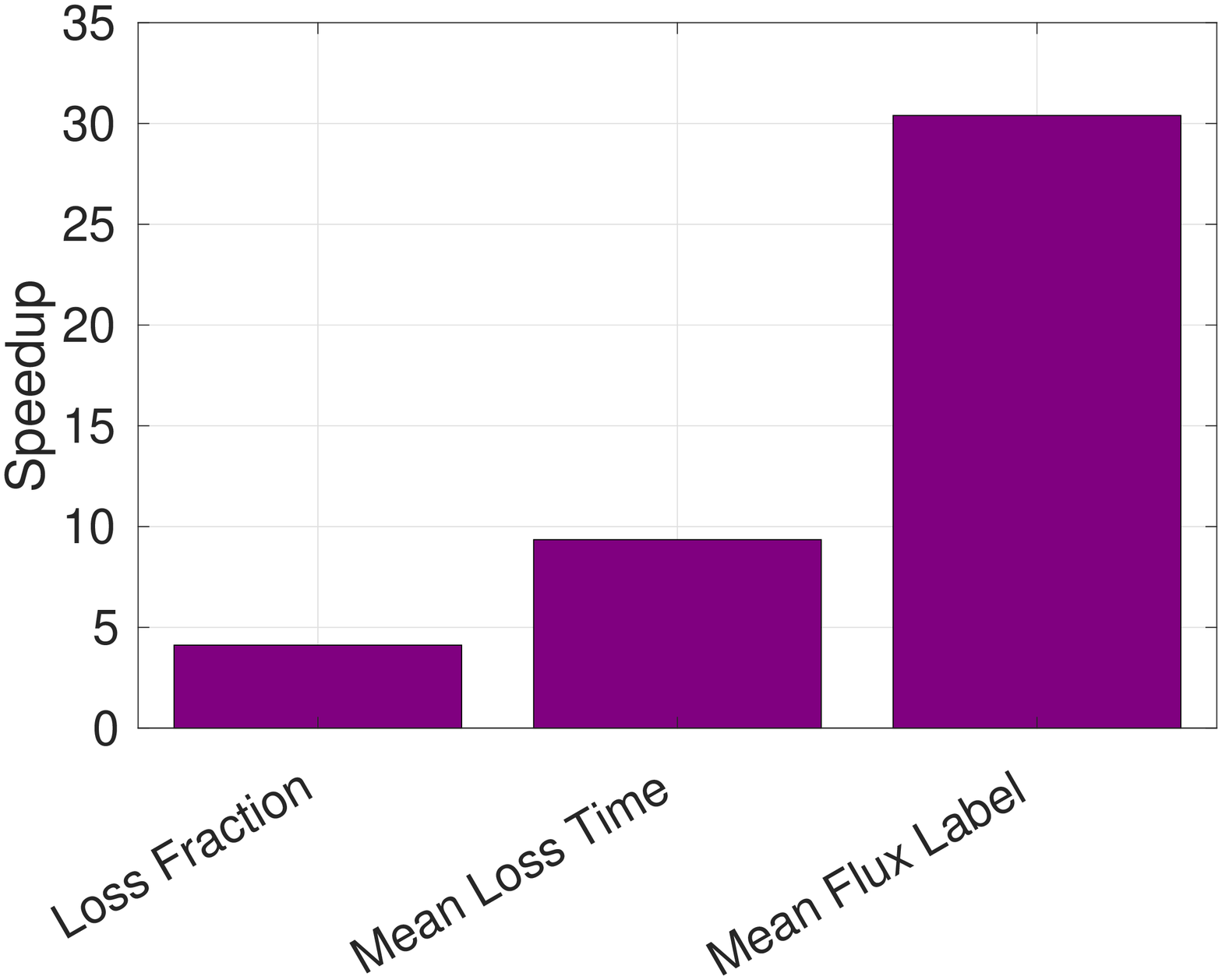}
    \caption{RMSE vs computational budget (seconds) for \textbf{Top left:} loss fraction \textbf{Top right:} mean modified loss time. \textbf{Bottom left:} mean normalized flux label. \textbf{Bottom right:} Computational speedup of MFMC compared to MC for each of the quantities of interest. For the mean flux label, we report the median over speedups for each of the $N_{\text{flux}}$ specified time points.}
    \label{fig:short-time}
\end{figure}

For a given computational budget $p$, we would ideally want to use the optimal MFMC estimator with $n$ high-fidelity evaluations and $m$ low-fidelity evaluations satisfying (\ref{optimal-MFMC-params}). However, in practice the resulting $n$ and $m$ are not necessarily integer. Thus we simply round down and choose
\begin{align*}
    n = \left\lfloor \frac{p}{1 + \sqrt{\frac{\rho(F,G)^{2}}{w(1-\rho(F,G)^{2})}}} \right\rfloor, \quad m = (p-n)w
\end{align*}
and compare this MFMC estimator to the MC estimator with $p$ high-fidelity model evaluations. In the case of the normalized flux label, we average the RMSE over the $N_{\text{flux}}$ specified time points.

In Figure \ref{fig:short-time} we see the clear variance reduction using MFMC with our low-fidelity models compared to standard Monte Carlo. We see that for the same computational budget, for each quantity of interest, MFMC is able to noticeably outperform MC. The most remarkable improvement is for the mean normalized flux label, where MFMC is nearly 30 times faster than MC. The improvement for the loss fraction is more modest, but MFMC is still 4 times faster than MC. 

\begin{figure}[b]
    \centering        \includegraphics[scale=0.345]{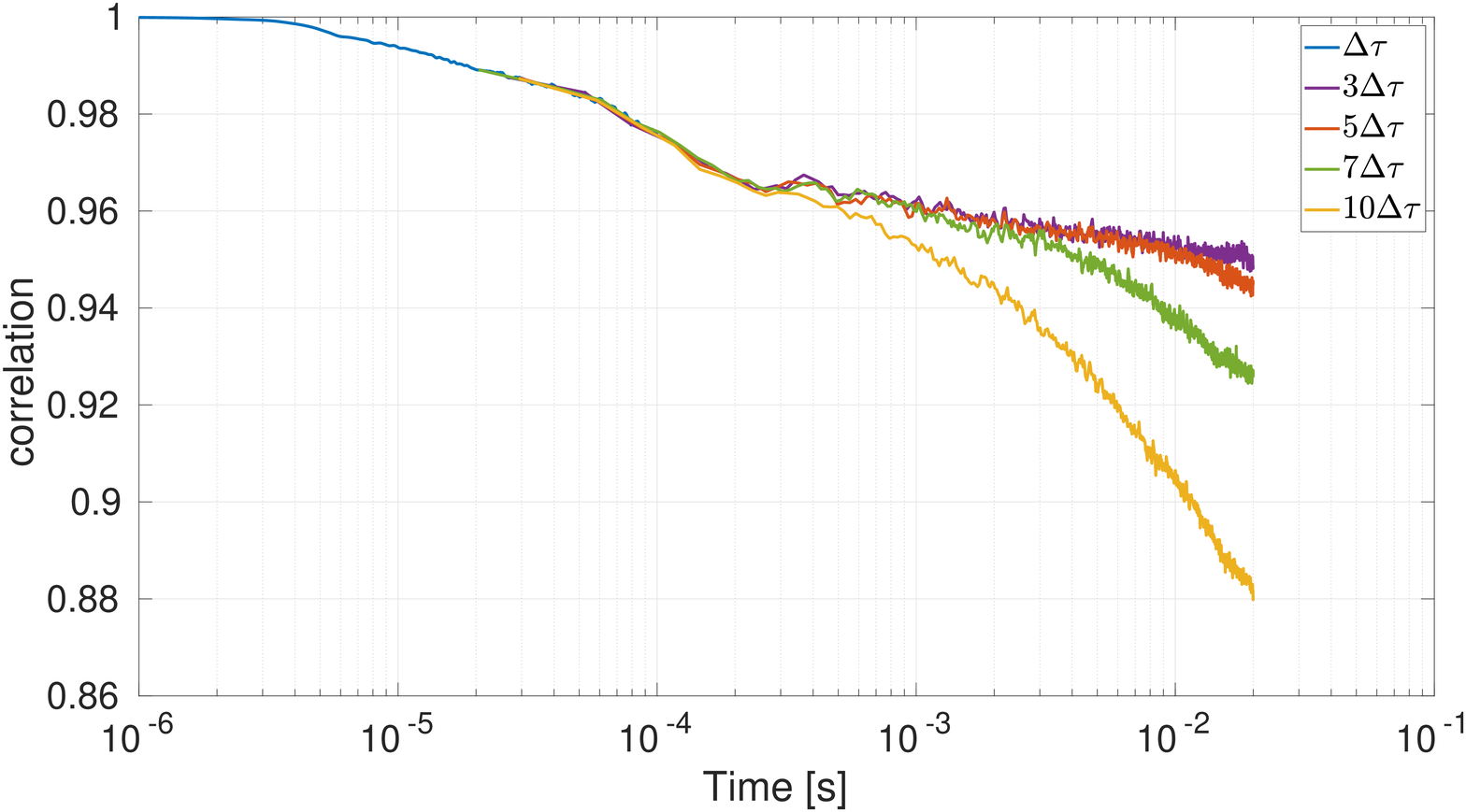}
    \caption{Correlation between $F^{\text{flux}}$ and $G^{\text{flux}}$ as a function of time, using different time step sizes for $F^{\text{flux}}$ (which in turn leads to a different $G^{\text{flux}}$ for each step size). We note that all the correlations stay practically the same until around $3 \times 10^{-4}$ seconds. After that, the experiment with the lowest resolution high-fidelity model (i.e. $10 \Delta \tau$) has a rapid drop off in correlation. The decrease with respect to different time step sizes is expected, with the highest resolution experiment maintaining the highest correlation.}
    \label{fig:flux-correlation-stepsize}
\end{figure}

In the case of the normalized flux label, each of the $N_{\text{flux}}$ specified time points has a different speedup, and in Fig. \ref{fig:short-time} we report the median over all such values. The different speedups at each specified time point $t_{k}$ is due to the fact that the correlation between $F^{\text{flux}}$ and $G^{\text{flux}}$ decreases over time, which can be seen in Fig. \ref{fig:flux-correlation-stepsize}. As a result the measured speedup for computing $\E[F^{\text{flux}}(t_{1}, \vb{R}_{0}, V_{0})]$ is much larger than the measured speedup of, e.g., $\E[F^{\text{flux}}(T_{\text{final}}, \vb{R}_{0}, V_{0})]$. Consequently, in utilizing the speedups for different time points we must be careful of outliers. This is the primary reasoning behind assessing the speedup for the whole trajectory by the median of the speedups at each time point, since the median is more robust to outliers than other measures of central tendency, such as the mean.

\subsection{Lower resolution high-fidelity model and longer time integration}

We now extend our analysis to a longer $T_{\text{final}}$ of 20ms. Since this is 200 times longer than the short time integration case, we must increase the step size for the high-fidelity model by a factor of 3, 5, 7, and 10 for computational tractability. Note that changing the step size changes the high-fidelity models $F^{\text{lost}}$, $F^{\text{time}}$, and $F^{\text{flux}}$, which implicitly changes the corresponding low-fidelity models $G^{\text{lost}}$, $G^{\text{time}}$, and $G^{\text{flux}}$.

The results in Table \ref{tab:correlation} indicate that accuracy of the high-fidelity model has a direct impact on the level of correlation between the high-fidelity and the low-fidelity models. The numerical experiments with $3 \Delta \tau$ and $5\Delta \tau$ yield similar correlations by the final time, but we see that as the time step of the high-fidelity numerical integrator further increases, and the accuracy of the high-fidelity model thus further decreases, the correlation with the low fidelity model decays quickly. A partial exception to that conclusion applies to the modified loss time, for which the correlation remains steady as we increase the time step of the ODE integrator. For the moment, we do not have an explanation for this surprising observation.
 
The importance of relying on an accurate high-fidelity model can be seen in Figure \ref{fig:flux-correlation-stepsize} as well, where we plot the correlation between $F^{\text{flux}}$ and $G^{\text{flux}}$ as a function of time, for different choices of step size. For short time all step sizes of the numerical ODE integrator provide the same correlation, but as time increases, we see that low resolution in the high-fidelity model leads to rapid loss of correlation. Moreover we notice the general pattern that, uniformly in time, the more resolution one has for the high-fidelity model, the more correlation one obtains with the corresponding low fidelity model. 

Since the costs of our low-fidelity models remain the same, it is doubly beneficial for MFMC that correlation between the high-fidelity and low-fidelity models tends to increase as the resolution, and thus cost, of the high-fidelity solver increases. Indeed, improving the resolution of the high-fidelity model not only yields better correlation to the resulting low-fidelity model, but also a larger cost ratio, which in turn means more variance reduction with MFMC.

\begin{figure}[t]
    \centering
    \includegraphics[scale=0.31]{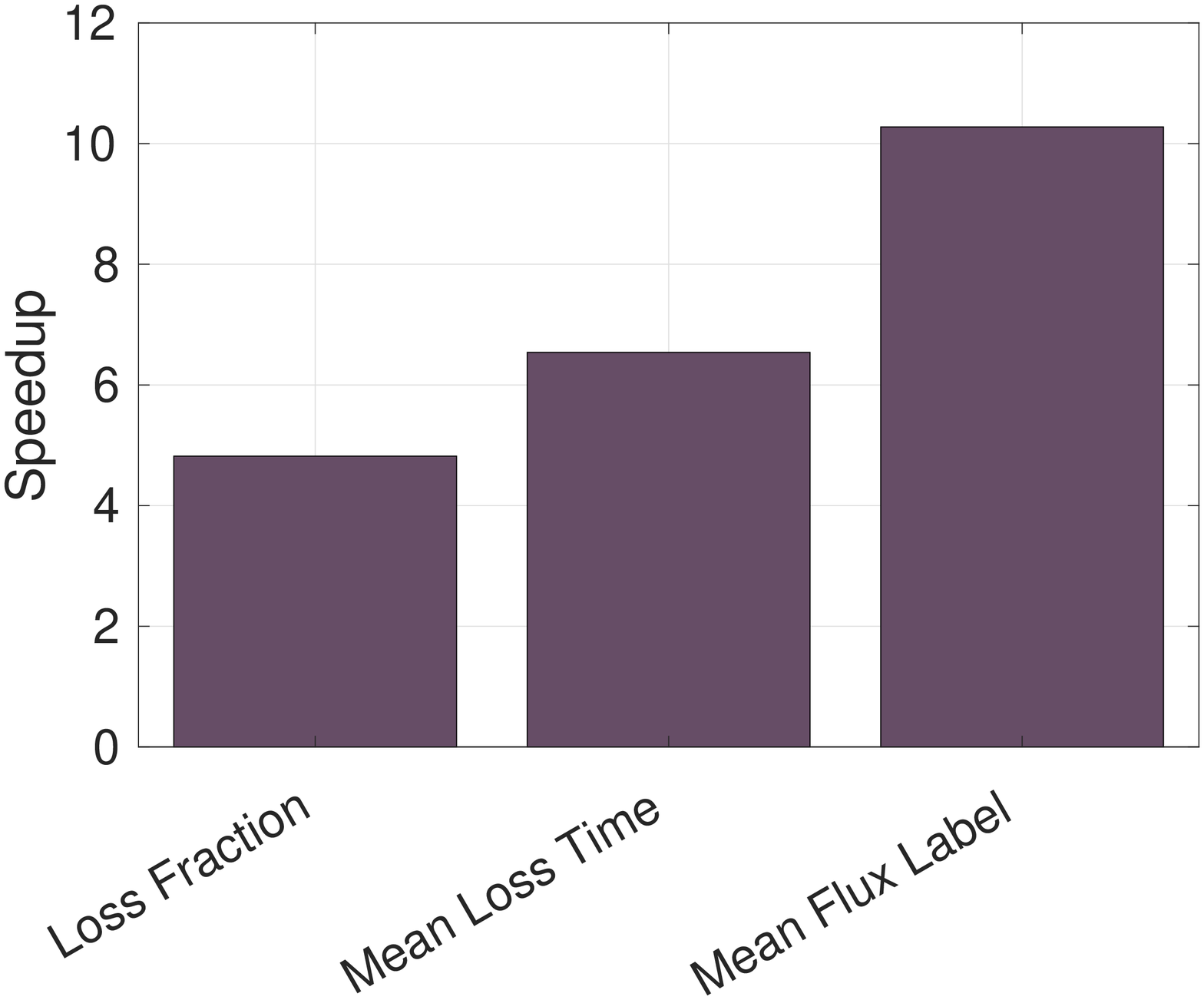}
    \includegraphics[scale=0.31]{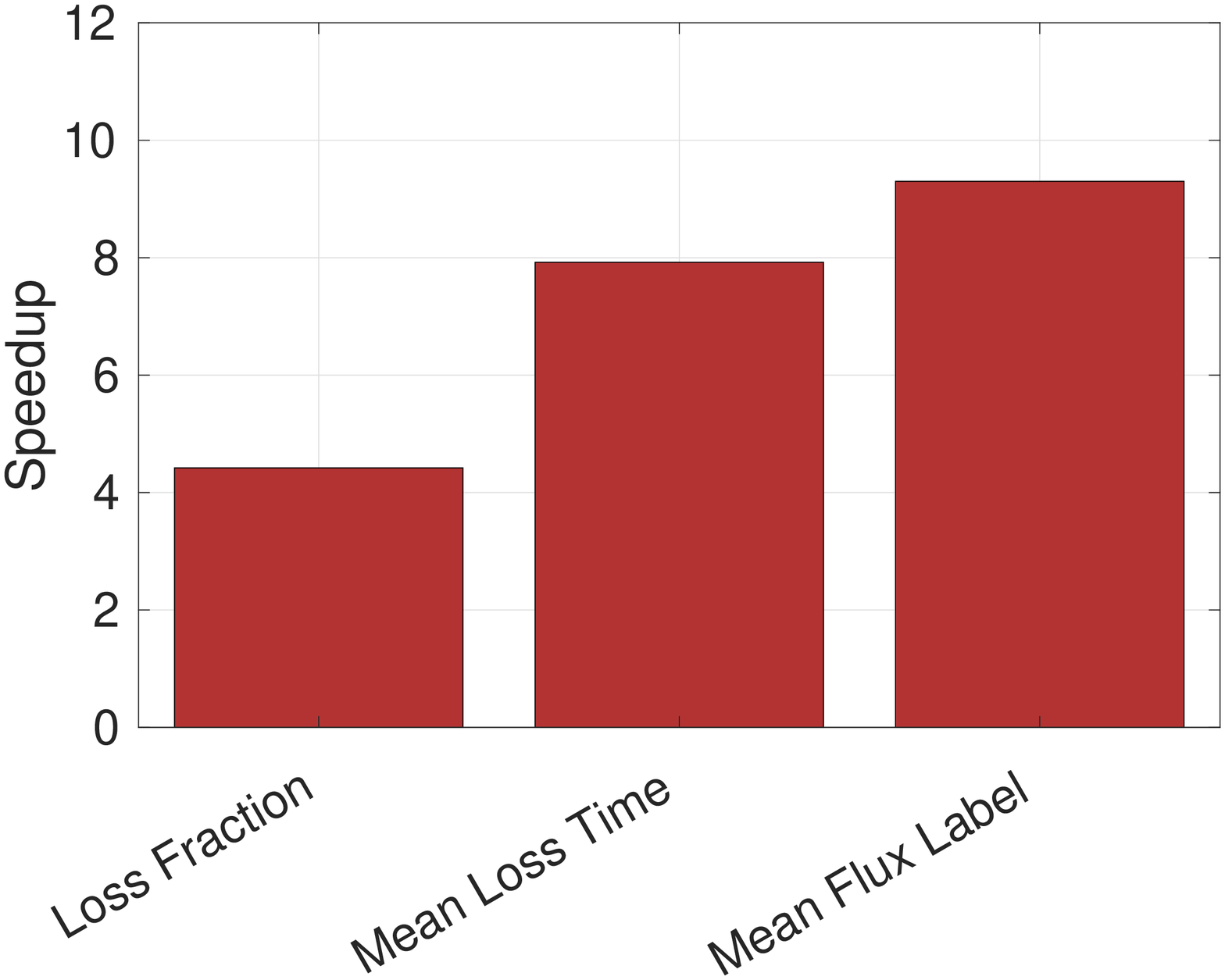}\\
    \includegraphics[scale=0.31]{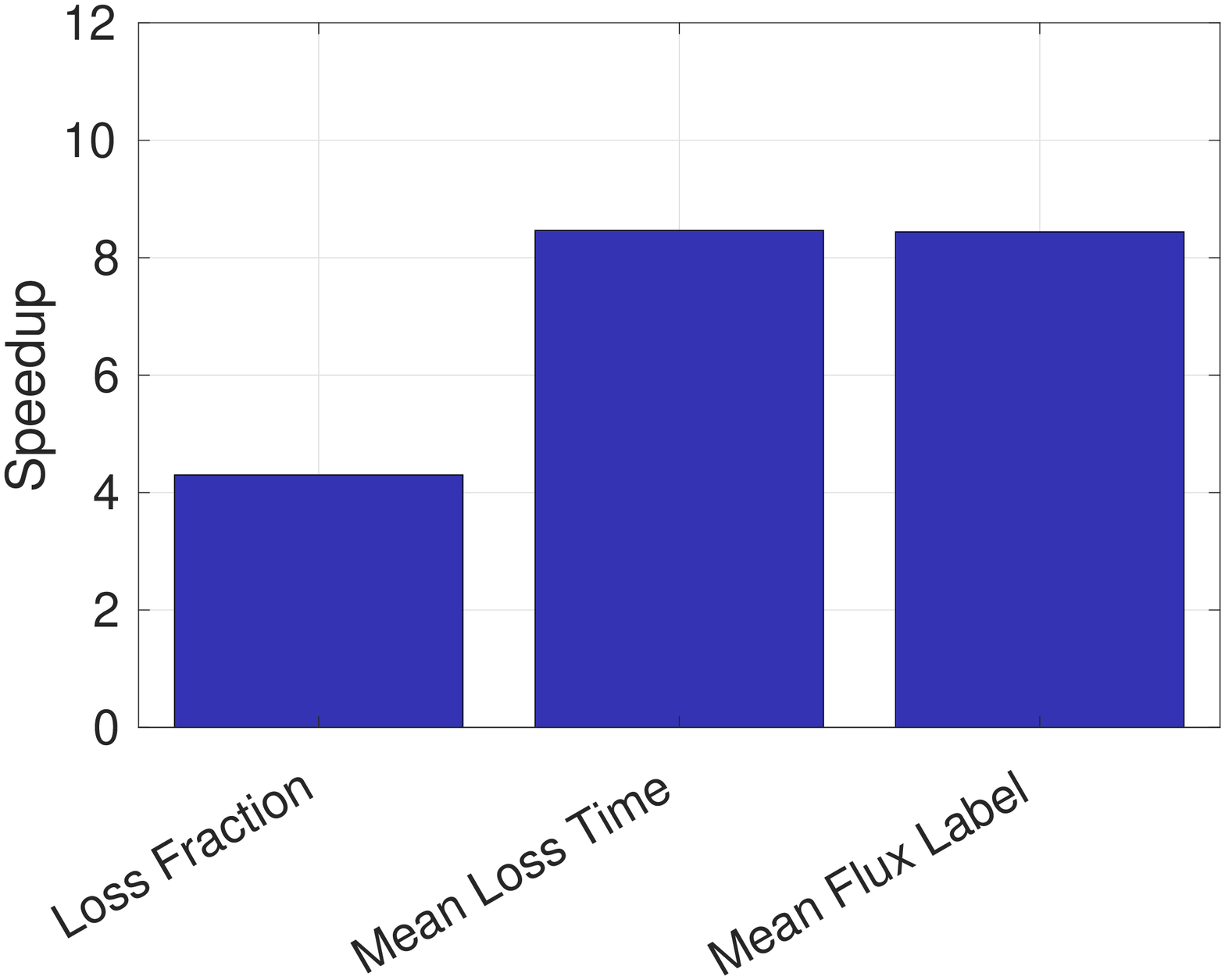}
    \includegraphics[scale=0.31]{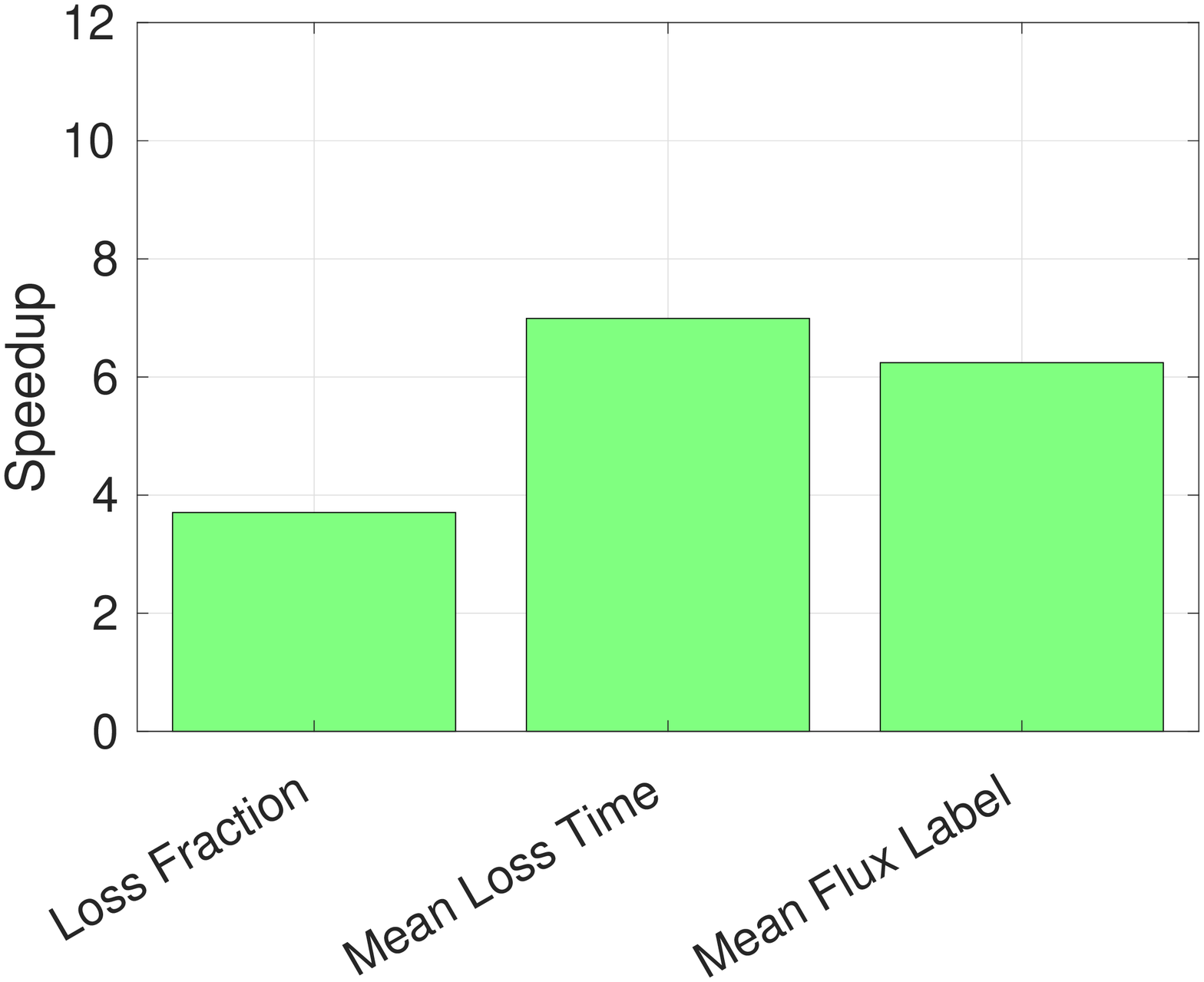}
    \caption{Measured speedup for $T_{\text{final}} = 0.02$s for different time step sizes of the high-fidelity ODE integrator, using $\Delta \tau = 2\pi / \Omega$. \textbf{Top Left}: $3\Delta \tau$. \textbf{Top Right}: $5\Delta \tau$. \textbf{Bottom Left}: $7\Delta \tau$. \textbf{Bottom Right}: $10 \Delta \tau$.}
    \label{fig:flux-longtime}
\end{figure}

    Just by looking at Tables \ref{tab:costs} and \ref{tab:correlation} and Figure \ref{fig:flux-correlation-stepsize}, we might expect some of the speedup we obtained in our high resolution, short time integration analysis to extend to the current analysis with lower resolution and longer time integration. Although the correlation goes down for longer time, the cost ratio grows rapidly, which in the MFMC setting can offset the effects of decreasing correlation. Furthermore, we may expect that the models with smaller step sizes (e.g. $3\Delta \tau$) will yield better speedup than models with larger step sizes (e.g. $10 \Delta \tau$). This is precisely what we now test numerically. 
    
    For computational budgets $p=100,250,500,1000$, we measure the RMSE of MC and MFMC by using 90 replicates for each estimator and compare them to a reference solution generated using MC with $10^{5}$ samples. We rely on fewer replicates in estimating the RMSE and have less resolution in the reference solution because of the significantly increased computational cost for longer time integration of the particle orbits. We balance the costs in the same manner as in the short time integration case, and also assess the flux label speedup using the median, for the same reasons as before.
    
    Our speedup results for the lower resolution, longer time integration experiments are displayed in Figure \ref{fig:flux-longtime}. Broadly, we see that we lose some speedup for this larger $T_{\text{final}}$ as compared to our short time integration experiment, but still achieve an order of magnitude speedup in estimating the mean flux label when the step size $3\Delta \tau$ is used. For the other quantities of interest, the speed up is also quite large, the smallest being an almost 4 times speedup for estimating the loss fraction with the high fidelity integrator with time step $10\Delta\tau$. For all time step sizes, the speed up for estimating the loss fraction is about 4, as we had obtained for the experiment with a short time integration.
    
    Comparing across experiments with different step sizes, we note that we typically get the most speedup for the most accurate, and most expensive, high-fidelity model, with step size $3 \Delta \tau$. As the step size increases, we see that the speedup in estimating the loss fraction decreases the slowest, whereas the speedup for estimating the mean flux label decreases the fastest. This is consistent with the results from Figure \ref{fig:flux-correlation-stepsize}, and it means that when we increase the step size, the larger cost ratio does not seem to fully make up for the decreased correlation. The speedup for the mean modified loss time does not appear to depend on the step size in a clear manner, and is between around 6.5 and 8 depending on the step size. These results are also consistent with the correlations reported in Table \ref{tab:correlation}, where we did not observe any obvious trend with the step size for this quantity of interest. Note that for the largest step size $10 \Delta \tau$, there is more speedup in estimating the mean modified loss time than in estimating the mean flux label.

\section{Summary and Discussion} \label{section:conclusion}

We have developed and implemented the first MFMC scheme for estimating energetic particle confinement in stellarator configurations. For the Wistell-A magnetic equilibrium we considered, this has allowed us to gain over an order of magnitude in speedup compared to standard Monte Carlo for certain cases and certain physical quantities. For other cases and physical quantities for which MFMC acceleration is less efficient, the speedup was still at least a factor of four. This speedup was gained by leveraging efficient data-driven surrogate models built from interpolants. 

Although we found that MFMC provided the largest speedup for our highest resolution high fidelity model with particle trajectories which were integrated over a short time scale, we observed that the speedups for estimating the loss fraction and the modified mean loss time remained large and roughly constant as we significantly increased the final time for the integration of the particle trajectories and reduced the resolution of our high-fidelity models.

As our results suggest, the acceleration obtained by replacing standard Monte Carlo with MFMC will depend on the physical quantity of interest, on the equilibrium magnetic configuration, and on the manner the uncertainty in the initial conditions is introduced. For example, one can mathematically show that speedup for the loss fraction deteriorates as the true loss fraction $\E[F^{\text{lost}}]$ goes to 0. If we had not assumed that alpha particles are born uniformly in the stellarator volume $U$, but instead had had all of them born on a flux surface close to the magnetic axis, as is sometimes done in confinement studies, the true loss fraction would have been smaller, and the MFMC speedup for this quantity of interest may have been smaller as well. On the other hand, the Wistell-A configuration is already well optimized for alpha particle confinement. If we had considered a magnetic configuration with significantly worse alpha particle confinement, we may have obtained a much higher speedup through our MFMC implementation. In the context of stellarator optimization, the approach we presented in this article may therefore be most useful in the early design optimization phase, when the loss fraction is high. In the later design phases, if the loss fraction has been optimized to become very low, the variance reduction provided by MFMC via control variates becomes ineffective for the estimation of the loss fraction (but not necessarily for other confinement quantities, as we showed in this work). For such situations, variance reduction methods based on, e.g., multi-fidelity importance sampling are more suitable \cite{Peherstorfer:2018}. We leave this for future work.

We applied our approach to a relatively straightforward setup, but our MFMC framework for estimating energetic particle confinement lends itself to extensions to more detailed physics and design studies. For example, it would be straightforward to replace our numerical ODE integrator with existing high performance solvers for guiding center trajectories, which may also include collisions \cite{Drevlak_2014,Pfefferle_2016} and wave-particle interactions \cite{Sigmar1992}. Another immediate extension would be to consider more sophisticated models for the initial conditions of the alpha particles, accounting for energy spread at birth, and the fact that for peaked pressure profiles, most alpha particles are born in the core of the device. Our success with MFMC using guiding center orbits also suggests that estimating statistics based on full-orbit trajectories may become computationally tractable. While full-orbit simulations are extremely expensive to perform, if we can find a proper surrogate to leverage, MFMC would enable high accuracy estimation of full-orbit statistics while only requiring a very small amount of full-orbit simulations. These topics are all subjects of ongoing work. Finally, it is important to stress that for a reliable analysis of energetic particle confinement, single-particle codes and theories may not be sufficient. A more self-consistent drift kinetic theory has recently been proposed for tokamaks with magnetic field perturbations \cite{tolman_catto_2021}, and we do not yet know if MFMC is relevant for such theory, and could lead to speed-up.

\section*{Acknowledgements}
We would like to thank Aaron Bader for providing the VMEC files for the Wistell-A magnetic equilibrium, and for many enlightening conversations on numerical methods for energetic particle confinement studies. Frederick Law was supported by the Department of Defense (DoD) through the National Defense Science \& Engineering Graduate (NDSEG) Fellowship Program and was supported in part by the Research Training Group in Modeling and Simulation funded by the National Science Foundation via grant RTG/DMS – 1646339. Antoine Cerfon was partially supported for this work by the Simons Foundation under grant No. 560651, and by the United States National Science Foundation under grant No. PHY-1820852. Benjamin Peherstorfer was partially supported by AFOSR under award number FA9550-21-1-0222 (Dr.~Fariba Fahroo) and by NSF under award number CMMI-1761068. 

\newpage
\bibliographystyle{plain}
\bibliography{refs.bib}

\begin{thebibliography}{10}

\bibitem{Albert:2020}
C.~G. Albert, S.~V. Kasilov, and W.~Kernbichler.
\newblock Accelerated methods for direct computation of fusion alpha particle
  losses within, stellarator optimization.
\newblock {\em Journal of Plasma Physics}, 86(2):815860201, 2020.

\bibitem{Appelbe_2011}
B~Appelbe and J~Chittenden.
\newblock The production spectrum in fusion plasmas.
\newblock {\em Plasma Physics and Controlled Fusion}, 53(4):045002, feb 2011.

\bibitem{bader2021energetic}
A~Bader, D~T Anderson, M~Drevlak, B~J Faber, C~C Hegna, S~Henneberg,
  M~Landreman, J~C Schmitt, Y~Suzuki, and A~Ware.
\newblock Energetic particle transport in optimized stellarators, 2021.

\bibitem{Bader:2019}
A.~Bader, M.~Drevlak, D.~T. Anderson, B.~J. Faber, C.~C. Hegna, K.~M. Likin,
  J.~C. Schmitt, and J.~N. Talmadge.
\newblock Stellarator equilibria with reactor relevant energetic particle
  losses.
\newblock {\em Journal of Plasma Physics}, 85(5):905850508, 2019.

\bibitem{Bader2020new}
A~Bader, BJ~Faber, JC~Schmitt, DT~Anderson, M~Drevlak, JM~Duff, H~Frerichs,
  CC~Hegna, TG~Kruger, M~Landreman, et~al.
\newblock A new optimized quasihelically symmetricstellarator.
\newblock {\em arXiv preprint arXiv:2004.11426}, 2020.

\bibitem{Beidler2001}
C.~D. Beidler, Ya.~I. Kolesnichenko, V.~S. Marchenko, I.~N. Sidorenko, and
  H.~Wobig.
\newblock Stochastic diffusion of energetic ions in optimized stellarators.
\newblock {\em Physics of Plasmas}, 8(6):2731--2738, 2001.

\bibitem{Boozer:1980}
Allen~H. Boozer.
\newblock Guiding center drift equations.
\newblock {\em The Physics of Fluids}, 23(5):904--908, 1980.

\bibitem{Brysk_1973}
H~Brysk.
\newblock Fusion neutron energies and spectra.
\newblock {\em Plasma Physics}, 15(7):611--617, jul 1973.

\bibitem{Bunno:2013}
M.~Bunno, Y.~Nakamura, Y.~Suzuki, K.~Shinohara, G.~Matsunaga, and K.~Tani.
\newblock Fusion alpha-particle losses in a high-beta rippled tokamak.
\newblock {\em Physics of Plasmas}, 20(8):082511, 2013.

\bibitem{BurbySenguptaKineticMHD}
J.~W. Burby and W.~Sengupta.
\newblock Hamiltonian structure of the guiding center plasma model.
\newblock {\em Physics of Plasmas}, 25(2):020703, 2018.

\bibitem{Cary:2009}
John~R. Cary and Alain~J. Brizard.
\newblock Hamiltonian theory of guiding-center motion.
\newblock {\em Rev. Mod. Phys.}, 81:693--738, May 2009.

\bibitem{CerfonFreidbergKineticMHD}
Antoine~J. Cerfon and Jeffrey~P. Freidberg.
\newblock Magnetohydrodynamic stability comparison theorems revisited.
\newblock {\em Physics of Plasmas}, 18(1):012505, 2011.

\bibitem{Cliffe:2011}
K.~A. Cliffe, M.~B. Giles, R.~Scheichl, and A.~L. Teckentrup.
\newblock Multilevel monte carlo methods and applications to elliptic pdes with
  random coefficients.
\newblock {\em Computing and Visualization in Science}, 14(3):3--15, 2011.

\bibitem{ColeCollisionlessXGC}
M.~D.~J. Cole, R.~Hager, T.~Moritaka, S.~Lazerson, R.~Kleiber, S.~Ku, and C.~S.
  Chang.
\newblock Comparative collisionless alpha particle confinement in stellarator
  reactors with the xgc gyrokinetic code.
\newblock {\em Physics of Plasmas}, 26(3):032506, 2019.

\bibitem{DiMarco2019}
Giacomo Dimarco and Lorenzo Pareschi.
\newblock Multi-scale control variate methods for uncertainty quantification in
  kinetic equations.
\newblock {\em Journal of Computational Physics}, 388:63--89, 2019.

\bibitem{DiMarco2020}
Giacomo Dimarco and Lorenzo Pareschi.
\newblock Multiscale variance reduction methods based on multiple control
  variates for kinetic equations with uncertainties.
\newblock {\em Multiscale Modeling \& Simulation}, 18(1):351--382, 2020.

\bibitem{Drevlak_2018}
M.~Drevlak, C.D. Beidler, J.~Geiger, P.~Helander, and Y.~Turkin.
\newblock Optimisation of stellarator equilibria with {ROSE}.
\newblock {\em Nuclear Fusion}, 59(1):016010, nov 2018.

\bibitem{Drevlak_2014}
M.~Drevlak, J.~Geiger, P.~Helander, and Y.~Turkin.
\newblock Fast particle confinement with optimized coil currents in the w7-x
  stellarator.
\newblock {\em Nuclear Fusion}, 54(7):073002, apr 2014.

\bibitem{E2009}
Weinan E, Weiqing Ren, and Eric Vanden-Eijnden.
\newblock A general strategy for designing seamless multiscale methods.
\newblock {\em Journal of Computational Physics}, 228(15):5437--5453, 2009.

\bibitem{Fatkullin2004}
Ibrahim Fatkullin and Eric Vanden-Eijnden.
\newblock A computational strategy for multiscale systems with applications to
  lorenz 96 model.
\newblock {\em Journal of Computational Physics}, 200(2):605--638, 2004.

\bibitem{GradGuidingCenter}
Harold Grad.
\newblock Variational principle for a guiding‐center plasma.
\newblock {\em The Physics of Fluids}, 9(2):225--251, 1966.

\bibitem{Grieger:1992}
G.~Grieger, W.~Lotz, P.~Merkel, J.~Nührenberg, J.~Sapper, E.~Strumberger,
  H.~Wobig, R.~Burhenn, V.~Erckmann, U.~Gasparino, L.~Giannone, H.~J. Hartfuss,
  R.~Jaenicke, G.~Kühner, H.~Ringler, A.~Weller, and F.~Wagner.
\newblock Physics optimization of stellarators.
\newblock {\em Physics of Fluids B: Plasma Physics}, 4(7):2081--2091, 1992.

\bibitem{MCMethods}
J.~M. Hammersley and D.~C. Handscomb.
\newblock {\em Monte Carlo Methods}.
\newblock Methuen London, 1964.

\bibitem{waelbroeck2018framework}
Richard~D. Hazeltine and Fran{\c{c}}ois~L Waelbroeck.
\newblock {\em The Framework of Plasma Physics}.
\newblock CRC Press, 2018.

\bibitem{Heidbrink_1994}
WW~Heidbrink and GJ~Sadler.
\newblock The behaviour of fast ions in tokamak experiments.
\newblock {\em Nuclear Fusion}, 34(4):535, 1994.

\bibitem{Helander_2014}
Per Helander.
\newblock Theory of plasma confinement in non-axisymmetric magnetic fields.
\newblock {\em Reports on Progress in Physics}, 77(8):087001, jul 2014.

\bibitem{Henneberg_2019}
S.A. Henneberg, M.~Drevlak, C.~Nührenberg, C.D. Beidler, Y.~Turkin, J.~Loizu,
  and P.~Helander.
\newblock Properties of a new quasi-axisymmetric configuration.
\newblock {\em Nuclear Fusion}, 59(2):026014, jan 2019.

\bibitem{Hirshman:1983}
S.~P. Hirshman and J.~C. Whitson.
\newblock Steepest‐descent moment method for three‐dimensional
  magnetohydrodynamic equilibria.
\newblock {\em The Physics of Fluids}, 26(12):3553--3568, 1983.

\bibitem{konrad2021}
Julia Konrad, Ionut-Gabriel Farcas, Benjamin Peherstorfer, Alessandro~Di Siena,
  Frank Jenko, Tobias Neckel, and Hans-Joachim Bungartz.
\newblock Data-driven low-fidelity models for multi-fidelity monte carlo
  sampling in plasma micro-turbulence analysis, 2021.

\bibitem{KruskalOberman}
M.~D. Kruskal and C.~R. Oberman.
\newblock On the stability of plasma in static equilibrium.
\newblock {\em The Physics of Fluids}, 1(4):275--280, 1958.

\bibitem{Ku_2006}
L.~P. Ku and P.~R. Garabedian.
\newblock New classes of quasi-axisymmetric stellarator configurations.
\newblock {\em Fusion Science and Technology}, 50(2):207--215, 2006.

\bibitem{Ku2008}
L.~P. Ku, P.~R. Garabedian, J.~Lyon, A.~Turnbull, A.~Grossman, T.~K. Mau,
  M.~Zarnstorff, and ARIES Team.
\newblock Physics design for aries-cs.
\newblock {\em Fusion Science and Technology}, 54(3):673--693, 2008.

\bibitem{Landreman:2018}
M.~Landreman.
\newblock {\em Guiding center drift in general flux coordinates}, 2018.

\bibitem{Littlejohn:1983}
R.~G. Littlejohn.
\newblock Variational principles of guiding centre motion.
\newblock {\em Journal of Plasma Physics}, 29(1):111–125, 1983.

\bibitem{Lotz_1992}
W~Lotz, P~Merkel, J~Nuhrenberg, and E~Strumberger.
\newblock Collisionless alpha -particle confinement in stellarators.
\newblock {\em Plasma Physics and Controlled Fusion}, 34(6):1037--1052, jun
  1992.

\bibitem{Merkel:1987}
P.~Merkel.
\newblock Solution of stellarator boundary value problems with external
  currents.
\newblock {\em Nuclear Fusion}, 27(5):867--871, 1987.

\bibitem{Mynick2006}
H.~E. Mynick, A.~H. Boozer, and L.~P. Ku.
\newblock Improving confinement in quasi-axisymmetric stellarators.
\newblock {\em Physics of Plasmas}, 13(6):064505, 2006.

\bibitem{Najmabadi_2008}
F.~Najmabadi, A.~R. Raffray, S.~I. Abdel-Khalik, L.~Bromberg, L.~Crosatti,
  L.~El-Guebaly, P.~R. Garabedian, A.~A. Grossman, D.~Henderson, A.~Ibrahim,
  T.~Ihli, T.~B. Kaiser, B.~Kiedrowski, L.~P. Ku, J.~F. Lyon, R.~Maingi,
  S.~Malang, C.~Martin, T.~K. Mau, B.~Merrill, R.~L. Moore, R.~J.~Peipert Jr.,
  D.~A. Petti, D.~L. Sadowski, M.~Sawan, J.~H. Schultz, R.~Slaybaugh, K.~T.
  Slattery, G.~Sviatoslavsky, A.~Turnbull, L.~M. Waganer, X.~R. Wang, J.~B.
  Weathers, P.~Wilson, J.~C.~Waldrop III, M.~Yoda, and M.~Zarnstorffh.
\newblock The aries-cs compact stellarator fusion power plant.
\newblock {\em Fusion Science and Technology}, 54(3):655--672, 2008.

\bibitem{nelson_control_1987}
Barry~L. Nelson.
\newblock On control variate estimators.
\newblock {\em Computers \& Operations Research}, 14(3):219--225, 1987.

\bibitem{Nemov_2014}
V.~V. Nemov, S.~V. Kasilov, and W.~Kernbichler.
\newblock Collisionless high energy particle losses in optimized stellarators
  calculated in real-space coordinates.
\newblock {\em Physics of Plasmas}, 21(6):062501, 2014.

\bibitem{Nemov_1999}
V.~V. Nemov, S.~V. Kasilov, W.~Kernbichler, and M.~F. Heyn.
\newblock Evaluation of {$1/\nu$} neoclassical transport in stellarators.
\newblock {\em Physics of Plasmas}, 6(12):4622--4632, 1999.

\bibitem{Nemov_2008}
V.~V. Nemov, S.~V. Kasilov, W.~Kernbichler, and G.~O. Leitold.
\newblock Poloidal motion of trapped particle orbits in real-space coordinates.
\newblock {\em Physics of Plasmas}, 15(5):052501, 2008.

\bibitem{Ng:2014}
Leo W.~T. Ng and Karen~E. Willcox.
\newblock Multifidelity approaches for optimization under uncertainty.
\newblock {\em International Journal for Numerical Methods in Engineering},
  100(10):746--772, 2014.

\bibitem{OhlbergerSuccess}
Mario Ohlberger and Stephan Rave.
\newblock {\em Reduced basis methods: Success, limitations and future
  challenges}, pages 1--12.
\newblock Publishing House of Slovak University of Technology in Bratislava,
  2016.

\bibitem{Paul_2020}
Elizabeth~J. Paul, Thomas Antonsen, Matt Landreman, and W.~Anthony Cooper.
\newblock Adjoint approach to calculating shape gradients for three-dimensional
  magnetic confinement equilibria. part 2. applications.
\newblock {\em Journal of Plasma Physics}, 86(1):905860103, 2020.

\bibitem{P19AMFMC}
B.~Peherstorfer.
\newblock Multifidelity monte carlo estimation with adaptive low-fidelity
  models.
\newblock {\em SIAM/ASA Journal on Uncertainty Quantification}, 7:579--603,
  2019.

\bibitem{Peherstorfer:2018}
B.~Peherstorfer, K.~Willcox, and M.~Gunzburger.
\newblock Survey of multifidelity methods in uncertainty propagation,
  inference, and optimization.
\newblock {\em SIAM Review}, 60(3):550--591, 2018.

\bibitem{Peherstorfer:2016}
Benjamin Peherstorfer, Karen Willcox, and Max Gunzburger.
\newblock Optimal model management for multifidelity monte carlo estimation.
\newblock {\em SIAM Journal on Scientific Computing}, 38(5):A3163--A3194, 2016.

\bibitem{Pfefferle_2016}
D.~Pfefferl{\'{e}}, W.A. Cooper, A.~Fasoli, and J.P. Graves.
\newblock Effects of magnetic ripple on 3d equilibrium and alpha particle
  confinement in the european {DEMO}.
\newblock {\em Nuclear Fusion}, 56(11):112002, jul 2016.

\bibitem{Ramos1}
J.~J. Ramos.
\newblock Quasineutrality and parallel force balance in kinetic
  magnetohydrodynamics.
\newblock {\em Journal of Plasma Physics}, 81(1):905810111, 2015.

\bibitem{Ramos2}
J.~J. Ramos.
\newblock On stability criteria for kinetic magnetohydrodynamics.
\newblock {\em Journal of Plasma Physics}, 82(6):905820607, 2016.

\bibitem{Reiman_1999}
A~Reiman, G~Fu, S~Hirshman, L~Ku, D~Monticello, H~Mynick, M~Redi, D~Spong,
  M~Zarnstorff, B~Blackwell, A~Boozer, A~Brooks, W~A Cooper, M~Drevlak,
  R~Goldston, J~Harris, M~Isaev, C~Kessel, Z~Lin, J~F Lyon, P~Merkel,
  M~Mikhailov, W~Miner, G~Neilson, M~Okamoto, N~Pomphrey, W~Reiersen,
  R~Sanchez, J~Schmidt, A~Subbotin, P~Valanju, K~Y Watanabe, R~White,
  N~Nakajima, and C~Nührenberg.
\newblock Physics design of a high-bbeta quasi-axisymmetric stellarator.
\newblock {\em Plasma Physics and Controlled Fusion}, 41(12B):B273--B283, dec
  1999.

\bibitem{Reiman:1986}
A.~Reiman and H.~Greenside.
\newblock Calculation of three-dimensional mhd equilibria with islands and
  stochastic regions.
\newblock {\em Computer Physics Communications}, 43(1):157--167, 1986.

\bibitem{RosenbluthRostokerKineticMHD}
M.~N. Rosenbluth and N.~Rostoker.
\newblock Theoretical structure of plasma equations.
\newblock {\em The Physics of Fluids}, 2(1):23--30, 1959.

\bibitem{Sigmar1992}
D.~J. Sigmar, C.~T. Hsu, R.~White, and C.~Z. Cheng.
\newblock Alpha‐particle losses from toroidicity‐induced alfvén
  eigenmodes. part ii: Monte carlo simulations and anomalous alpha‐loss
  processes.
\newblock {\em Physics of Fluids B: Plasma Physics}, 4(6):1506--1516, 1992.

\bibitem{spitzer1956physics}
L~Spitzer.
\newblock Physics of fully ionised gases interscience.
\newblock {\em New York}, 1956.

\bibitem{Spong_2011}
D.~A. Spong.
\newblock Three-dimensional effects on energetic particle confinement and
  stability.
\newblock {\em Physics of Plasmas}, 18(5):056109, 2011.

\bibitem{Spong_2015}
Donald~A. Spong.
\newblock 3d toroidal physics: Testing the boundaries of symmetry breaking.
\newblock {\em Physics of Plasmas}, 22(5):055602, 2015.

\bibitem{tolman_catto_2021}
Elizabeth~A. Tolman and Peter~J. Catto.
\newblock Drift kinetic theory of alpha transport by tokamak perturbations.
\newblock {\em Journal of Plasma Physics}, 87(2):855870201, 2021.

\bibitem{Lee2014}
Bjorn~Engquist Yoonsang~Lee.
\newblock Variable step size multiscale methods for stiff and highly
  oscillatory dynamical systems.
\newblock {\em Discrete \& Continuous Dynamical Systems}, 34(3):1079--1097,
  2014.

\bibitem{Zweben_1992}
S.J Zweben, G.W Hammett, R.L Boivin, C.K Phillips, and J.R Wilson.
\newblock Loss of {MeV} ions during3he minority ion cyclotron resonance heating
  in {TFTR}.
\newblock {\em Nuclear Fusion}, 32(10):1823--1833, oct 1992.

\end{thebibliography}

\end{document}